\documentclass[a4paper,11pt]{article}
\usepackage[labelfont=bf,labelsep=period]{caption}
\usepackage[nodayofweek]{datetime}
\usepackage{enumitem}
\usepackage{float}
\usepackage[margin=2.5cm]{geometry}
\usepackage{graphicx}
\usepackage{numbertabbing}
\usepackage{times}
\usepackage{url}
\usepackage{xspace}
\usepackage{lscape}
\usepackage{subcaption}
\usepackage{epstopdf}
\usepackage{hyperref}
\usepackage{xcolor}

\usepackage{amsmath} 
\allowdisplaybreaks[2]          
\usepackage{amssymb} 
\usepackage{amsthm} 
\newtheoremstyle{plain-boldhead}
  {\topsep}
  {\topsep}
  {\itshape}
  {}
  {\bfseries}
  {.}
  { }
  {\thmname{#1}\thmnumber{ #2}\thmnote{ (\bfseries #3)}}
\newtheoremstyle{definition-boldhead}
  {\topsep}
  {\topsep}
  {\normalfont}
  {}
  {\bfseries}
  {.}
  { }
  {\thmname{#1}\thmnumber{ #2}\thmnote{ (\bfseries #3)}}
\theoremstyle{plain-boldhead}

\theoremstyle{definition-boldhead}
\newtheorem{definition}{Definition}

\newdateformat{simple}{\THEDAY\ \monthname[\THEMONTH]\ \THEYEAR}
\simple

\floatstyle{ruled}
\newfloat{algo}{htbp}{algo}
\floatname{algo}{Algorithm}
\newfloat{module}{htbp}{module}
\floatname{module}{Module}

\def \ifempty#1{\def\temp{#1} \ifx\temp\empty }

\newcommand{\str}[1]{\textsc{#1}}
\newcommand{\var}[1]{\textit{#1}}
\newcommand{\op}[1]{\textsl{#1}}
\newcommand{\msg}[2]{\ensuremath{\ifempty{#2} [\str{#1}] \else [\str{#1}, {#2}] \fi}}

\newcommand{\false}{\textsc{false}\xspace}
\newcommand{\true}{\textsc{true}\xspace}
\newcommand{\etal}{\emph{et al.}}

\newcommand{\eventmodp}[3]{$\langle$~{\var{#1}}-\textsl{#2}~$\mid$~{#3}~$\rangle$}
\newcommand\emptyhashmap{[\,]}

\newcommand{\CP}{\ensuremath{\mathcal{P}}\xspace}

\newcommand{\tx}{\var{tx}}

\newcommand\vbc{\var{vbc}\xspace}

\newcommand\cview{\var{view}\xspace}
\newcommand\payload{\var{payload}\xspace}
\newcommand\block{\var{block}\xspace}

\newcommand\vote{\var{vote}\xspace}

\newcommand\qc{\var{qc}\xspace}
\newcommand\inround{\var{inround}\xspace}
\newcommand\rp{\var{roundp}\xspace}
\newcommand\rd{\var{roundd}\xspace}
\newcommand\cblocks{\var{commitedBlocks}\xspace}
\newcommand\cblock{\var{commitedBlock}\xspace}
\newcommand\votes{\var{votes}\xspace}
\newcommand\bid{\var{block.id}\xspace}

\newcommand\view{\var{view}\xspace}
\newcommand\pb{\var{parentBlock}\xspace}
\newcommand\gb{\var{grandparentBlock}\xspace}
\newcommand\txp{$\var{tx}^\prime$\xspace}
\newcommand\ver{\var{vertices}\xspace}
\newcommand\cmp{\var{components}\xspace}
\newcommand\visited{\var{visited}\xspace}
\newcommand\stack{\var{stack}\xspace}
\newcommand\node{\var{node}\xspace}
\newcommand\transposed{\var{transposed}\xspace}
\newcommand\neighbour{\var{neighbour}\xspace}
\newcommand\vc{\var{vc}\xspace}
\newcommand\msgs{\var{msgs}\xspace}

\begin{document}

\title{Quick Order Fairness: Implementation and Evaluation}

\author{Christian Cachin$^1$\\
  University of Bern\\
  \url{christian.cachin@unibe.ch}
  \and Jovana Mićić$^1$\\
University of Bern\\
  \url{jovana.micic@unibe.ch}
}

\footnotetext[1]{Institute of Computer Science, University of Bern,
  Neubr\"{u}ckstrasse 10, 3012 CH-Bern, Switzerland.}

\maketitle

\begin{abstract}\noindent
  Decentralized finance revolutionizes traditional financial systems by leveraging blockchain technology to reduce trust. However, some vulnerabilities persist, notably \emph{front-running} by malicious actors who exploit transaction information to gain financial advantage. Consensus with a \emph{fair order} aims at preventing such attacks, and in particular, the \emph{differential order fairness} property addresses this problem and connects fair ordering to the validity of consensus. The notion is implemented by the Quick Order-Fair Atomic Broadcast (QOF) protocol (Cachin \etal, FC '22). This paper revisits the QOF protocol and describes a modular implementation that uses a generic consensus component. Moreover, an empirical evaluation is performed to compare the performance of QOF to a consensus protocol without fairness. Measurements show that the increased complexity comes at a cost, throughput decreases by at most 5\%, and latency increases by roughly 50ms, using an emulated ideal network. This paper contributes to a comprehensive understanding of practical aspects regarding differential order fairness with the QOF protocol and also connects this with similar fairness-imposing protocols like Themis and Pompē.
  \medskip
    

\end{abstract}

\section{Introduction}
\label{sec:intro}
\emph{Decentralized finance (DeFi)} describes a financial system built on blockchain technology that aims to recreate and enhance traditional financial services without relying on centralized authorities, such as banks or brokers. In the DeFi ecosystem, users can engage in various financial activities, including lending, borrowing, trading, and earning interest. DeFi mechanisms are implemented by smart contracts running on a decentralized blockchain network. Many parties jointly power the network through a robust consensus protocol against attacks by malicious actors.

DeFi, however, is not (yet) immune against certain kinds of fraud that have also been observed in the existing finance system. Traditional finance considers banks to be trusted entities, but malicious employees could gain an unfair advantage by exploiting privileged information in so-called insider trading. Consequently, regulation has been established that forbids such actions. Still, in DeFi, just by observing the freshly submitted transactions in the public network, a malicious party can use this information and manipulate the order of transactions in a block, leading to a \emph{front-running} attack. For example, this occurs when a malicious party exploits information about upcoming transactions to gain a financial profit, named \emph{maximal extractable value} (MEV) by Daian~\etal~\cite{DBLP:conf/sp/DaianGKLZBBJ20}. This is typically done with a \emph{sandwich attack} that strategically places two fraudulent transactions around a victim's transaction, one before and one after. Such attacks exploit the decentralized and transparent nature of the consensus and transaction execution in a blockchain, highlighting the need for a protocol that imposes fairness in DeFi. Moreover, front-running may occur also on non-programmable blockchains like the XRP Ledger~\cite{DBLP:conf/icbc2/TumasPTS23}.

In order to prevent such attacks, research has  proposed several solutions. One is to enforce a \emph{causal order} using distributed cryptography. A second solution, called \emph{receive-order fairness}, considers in which order transactions were received by the parties running the consensus protocol and enforces corresponding constraints on the order resulting from consensus. A third approach removes the influence on transaction order by the parties running consensus.  

\emph{Quick Order-Fair Atomic Broadcast (QOF)}~\cite{DBLP:conf/fc/CachinMSZ22} is a representative of the group of receive-order fairness protocols. It adds a new property called \emph{differential order fairness} to the existing properties of atomic broadcast. The protocol works for asynchronous and eventually synchronous networks with optimal resilience and tolerates faults of up to one-third of the total number of parties. The resilience to faults does not depend on the fairness notion. Compared to similar solutions, QOF is more efficient. It requires, on average, $O(n^2)$ messages to deliver one transaction. For comparison, the asynchronous \emph{Aequitas} protocol~\cite[Sec.~7]{DBLP:conf/crypto/Kelkar0GJ20} needs $O(n^4)$ messages and has resilience $n>4f$ or worse. Protocol \emph{Themis}~\cite{DBLP:conf/ccs/KelkarDLJK23} achieves the same resilience and takes also $O(n^2)$ messages to deliver one transaction, though a SNARK-based variant reduces this further in the best case. The number of faulty parties tolerated by Aequitas and Themis depends on the quality of the fairness that is achieved, however.

This paper revisits the order-fair atomic broadcast and the QOF protocol. Our primary motivation is to describe an implementation of the QOF protocol in detail and to measure its cost. Implementing a prototype is essential for empirically validating the theoretical base of the proposed solution. More precisely, we describe a modular implementation of the QOF protocol on top of an existing library for atomic (i.e., total-order) broadcast called \emph{bamboo}~\cite{DBLP:conf/icdcs/GaiFNFBD21}, which realizes the HotStuff~\cite{DBLP:conf/podc/YinMRGA19} consensus protocol (note that \emph{consensus} and \emph{atomic broadcast} are synonyms here). It uses three components: Byzantine consistent broadcast, validated Byzantine consensus, and a graph module. The first module provides consistency for transactions sent by potentially faulty parties~\cite{DBLP:conf/fc/CachinMSZ22}. The validated Byzantine consensus module is built directly on bamboo and supports \emph{external validity}. The graph module maintains a directed acyclic graph and provides functions for capturing potential dependencies among the transaction. Connecting all modules and implementing the logic of extracting the fair order of transactions from the graph structure completes the implementation of the QOF protocol. 

The implementation allows us to evaluate the performance of the QOF protocol and assess its efficiency. We measure throughput and latency and compare it to the baseline HotStuff implementation in bamboo~\cite{DBLP:conf/icdcs/GaiFNFBD21}. Our experiments indicate that (with four servers) compared to the HotStuff protocol, QOF reduces the throughput by at most 5\% and increases latency by about 50ms, reflecting the impact of the QOF protocol's increased complexity in an ideal, emulated network. We also draw a connection between our results and the performance of similar protocols for imposing a fair order, including Themis~\cite{DBLP:conf/ccs/KelkarDLJK23} and Pompē~\cite{DBLP:conf/osdi/ZhangSCZA20}.

Furthermore, we outline how the Quick Order-Fair protocol may be deployed in practice, offering two integration approaches. The first approach involves implementing it as a separate service, where clients submit transactions to ordering nodes. Ordering nodes use the QOF protocol to determine a fair order, which validators execute through a smart contract. The second approach directly integrates QOF into validators, with clients submitting transactions to validators running the algorithm and executing transactions on the ledger. 

To summarize, this paper presents three contributions:
\begin{itemize}
  \item It describes a practical implementation of the QOF protocol, illustrating many aspects left out in earlier work and providing a coherent representation of the protocol and its components.
  \item It examines the protocol's integration into real-world systems, explaining possible designs and providing a smart contract blueprint.
  \item It conducts an empirical evaluation in several dimensions: scalability, throughput, and latency. This evaluation affirms the efficacy of the QOF protocol and provides valuable guidance for its practical deployment in decentralized systems.
\end{itemize}

The paper starts with a review of the existing techniques for preventing MEV (Section~\ref{sec:related}). Then, we describe the Quick Order Fairness protocol (Section~\ref{sec:qof}). Section~\ref{sec:implementation} describes the prototype's building blocks and implementation. Section~\ref{sec:integration} describes how the protocol can be integrated into practical systems. Section~\ref{sec:evaluation} presents the evaluation results, and finally, Section~\ref{sec:conclusion} concludes the paper.

\section{Related work}
\label{sec:related}
Several lines of research have been developed in the past years to address the front-running problem at the consensus level of blockchain networks, both in academia and in practice. On a high level, we can group the proposed defense methods into three categories. Methods of the first kind aim to prevent side information leaks to malicious insiders by using distributed cryptography, which enforces a \emph{causal order} on the transaction sequence produced by consensus. The second kind of defense, known as \emph{receive-order fairness}, considers how the transactions were received by the individual parties running the consensus protocol and enforces corresponding constraints on the order resulting from consensus. The last category \emph{removes the influence on transaction order} by the parties running consensus. The following paragraphs present the most relevant work in each category.

\paragraph{Causal order.} The initial idea for outputting transactions correctly is to force the consensus protocol to respect a \emph{causal order} among the transactions according to how they were input, as defined by Reiter and Birman~\cite{DBLP:journals/toplas/ReiterB94}. They call this \emph{input causality}, a notion that has later been refined and made more formal by Cachin~\etal~\cite{DBLP:conf/crypto/CachinKPS01} as \emph{secure causal order}. It ensures that a given transaction $\tx$ must appear in the output sequence only after all transactions that potentially caused~$\tx$ have appeared, similar to notions of causality~\cite{hadtou93}. To ensure this, Reiter and Birman use threshold cryptography: A client encrypts its transaction with a public key held by the network, and the consensus protocol first decides on an order, subsequently decrypts the transactions, and executes them. Other approaches of this kind employ verifiable secret sharing (VSS)~\cite{DBLP:conf/dsn/DuanRZ17}, time-lock encryption~\cite{rivest1996time}, and delay encryption~\cite{DBLP:conf/crypto/BonehBBF18, DBLP:conf/eurocrypt/BurdgesF21, DBLP:conf/aft/ChiangDEG23}. For instance, \emph{Flash Freezing Flash Boys} (F3B) protocol~\cite{DBLP:conf/aft/ZhangMQBEF23} addresses front-running attacks by using threshold cryptography, and it is evaluated on the Ethereum network. Authors show that F3B can be deployed with a small overhear. 
\emph{Helix}~\cite{DBLP:journals/tnsm/YakiraACGLRT21}, commits to transaction ordering without seeing the content of a transaction. This is achieved using a commit-reveal protocol, which first commits the transactions and reveals the committed transactions only when the ordering has already been determined. Malkhi and Szalachowski~\cite{DBLP:conf/tokenomics/MalkhiS22} construct a protocol called \emph{Fino} which combines the Order-then-Reveal method on a directed acyclic graphs (DAG). 

\paragraph{Receive-order fairness.} A second line of research called \emph{receive-order fairness} stipulates that clients send their transactions simultaneously to all parties responsible for ordering. The consensus protocol considers the order of receiving transactions at a majority of the parties and respects that in the output. If enough parties receive some transaction \var{tx} before \txp, then \txp cannot be in the output before \var{tx}~\cite{DBLP:conf/crypto/Kelkar0GJ20}. Since every party has a local receive order, consensus must find one common order that matches the local observations and tolerates wrong reports by faulty parties. \emph{Order fairness} in this sense has been introduced by Kelkar~\etal~\cite{DBLP:conf/crypto/Kelkar0GJ20}. They also recognized that cycles may appear among the locally reported orders according to the \emph{Condorcet paradox} such that no output order exists that satisfies all input constraints. They introduced a notion called \emph{block order fairness}, implemented in Aequitas protocol, which outputs multiple transactions together in one ``block'' without deciding on an order within the block. Subsequent work by Kelkar~\etal~\cite{DBLP:conf/ccs/KelkarDLJK23} introduces a new technique called \emph{deferred ordering}, which overcomes a liveness issue in the Aequitas protocol and proposes a new protocol called Themis. Cachin~\etal~\cite{DBLP:conf/fc/CachinMSZ22} introduced a new notion of \emph{differential order fairness} and proposed a new protocol called \emph{Quick order-fair atomic broadcast} with an optimal resilience of $n>3f$. Section~\ref{sec:qof} presents more details about this protocol.

Order fairness has been defined differently by Kursawe~\cite{DBLP:conf/aft/Kursawe20} and Zhang~\etal~\cite{DBLP:conf/osdi/ZhangSCZA20}. These notions are known as \emph{timed relative fairness} and \emph{ordered linearizability}, respectively. Timed relative fairness is implemented by the Wendy~\cite{DBLP:conf/aft/Kursawe20} protocol, predicated on the idea that all parties have access to synchronized local clocks. For instance, if all correct parties observed that transaction $\tx$ needed to be ordered before $\tx'$, then $\tx$ would be scheduled and delivered before $\tx'$. Ordered linearizability assumes that the highest timestamp for transaction $\tx$ is smaller than the lowest timestamp for transaction $\tx$ in the output sequence and is implemented by the Pompē~\cite{DBLP:conf/osdi/ZhangSCZA20} protocol. For instance, in the output sequence, $\tx$ will show before $\tx'$ if the highest timestamp supplied by any correct party for a transaction $\tx$ is less than the lowest timestamp supplied by any correct party for a transaction $\tx'$. Based on a median calculation, ordered linearizability is implemented; however, as demonstrated by Kelkar~\etal~\cite{DBLP:conf/asiapkc/KelkarDK22}, this computation can be influenced by malicious parties.

\paragraph{Randomized order.}The blockchain community widely knows the concept of randomizing transaction order within a block. While Yanai~\cite{blinderswap} and others have explored and implemented this idea in practice, for instance, Randomspam~\cite{randomspam} recognizes the possibility of spamming attacks associated with randomized transactions. In such scenarios, attackers strategically insert numerous low-cost transactions to increase the likelihood of positioning some of these transactions precisely at a profitable transaction.
Another solution by Alpos~\etal~\cite{DBLP:journals/corr/abs-2307-02954} implement randomization using on-chain randomness, accompanied by a comprehensive security analysis. They proposed \emph{Partitioned and Permuted Protocol} ($\Pi^3$), in which the final order of transactions in the block is determined by a permutation that the miners generate. Leaders from the past blocks commit to their contributions to the permutation before the block is mined. Running an attack can sometimes be more profitable than the reward, so the authors proposed transaction chunking to enlarge the permutation space.
Non-programmable blockchains, like Ripple\footnote{\url{https://ripple.com/}},  also suffer from front-running attack~\cite{DBLP:conf/icbc2/TumasPTS23}. In the initial XRP Ledger version, the attacker could generate a lower transaction ID to ensure his transaction is executed before the victim's. Ripple's developers later fixed this problem by implementing a new transaction ordering strategy that creates a pseudo-random shuffle on the transaction ordering. A random salt derived from the accepted transactions ensures that all participants compute the same ordering. Tumas~\etal~\cite{DBLP:conf/icbc2/TumasPTS23} shows two strategies to perform the attack and concludes that the shuffling algorithm provides some security against naive attacks but does not prevent them. 
The concept of randomizing transactions is proposed for some other purposes as well. For instance, Chitra~\etal~\cite{DBLP:conf/fc/ChitraAE22} proposed a privacy-enchasing mechanism called \emph{Uniform Random Execution} that is used in constant function market makers (CFMMs) to provide the privacy for users. The proposed protocol splits large trades randomly and subsequently permutes the trade ordering. 

\paragraph{Overviews.} Finally, Heimbach and Wattenhofer~\cite{DBLP:conf/aft/HeimbachW22} give an overview of state-of-the-art techniques for preventing manipulation of the transaction order. They present a taxonomy of the techniques and compare them regarding decentralization, security, scope, and other properties. Another survey of knowledge given by Baum~\etal~\cite{DBLP:conf/fc/BaumCDFG22} describes common front-running attacks and assesses three mitigation categories. Moreover, they introduce a sandwich attack on input batching techniques that can be mitigated with private user balances and secret input stores. Both works conclude that despite the growing number of approaches to address transaction reordering manipulations on blockchains, an effective technique to mitigate the front-running problem still needs to be discovered. The current state of research indicates that no existing approach can fully meet the requirements posed by a decentralized blockchain environment.

\section{Quick Order Fairness Protocol}
\label{sec:qof}
This section reviews the Quick Order Fairness protocol (QOF)~\cite{DBLP:conf/fc/CachinMSZ22}. The protocol functions effectively in both asynchronous and eventually synchronous networks, demonstrating resilience by tolerating corruptions in up to a third of the parties. It is accessed with $\op{of-broadast(\tx)}$ for broadcasting a transaction $\tx$ and outputs transactions through $\op{of-deliver(T)}$, where $T$ is a set of transactions delivered at the same time. If one correct party \op{of-broadcasts} some transaction $\tx$, then every correct party eventually also \op{of-broadcasts} $\tx$. In order to implement \emph{$\kappa$-differentially order-fair atomic broadcast} correctly, QOF needs to satisfy the following properties:

\begin{definition}[$\kappa$-Differentially Order-Fair Atomic Broadcast] \label{def:diff-abc}
  A protocol for \emph{$\kappa$-differentially order-fair atomic broadcast}
  satisfies the properties \emph{no duplication, agreement} and \emph{total order} of \emph{atomic broadcast} and additionally:
  \begin{description}
  \item [Weak validity:] If all parties are correct and \op{of-broadcast}
    a finite number of transactions, then every correct party eventually
    \op{of-delivers} all of these \op{of-broadcast} transacations.
  \item [$\kappa$-differential order fairness:] If
    $b(\tx,\tx') > b(\tx',\tx) + 2f + \kappa$, then no correct party
    \op{of-delivers} $\tx'$ before $\tx$.
  \end{description}
\end{definition}

The value $b(\tx,\tx')$ denotes the \emph{number of correct parties} that \op{of-broadcast} $\tx$ before $\tx'$ in an execution. Parameter $f$ denotes the number of faulty parties, and fairness parameter $\kappa \geq 0$ expresses the strength of the fairness. Smaller values of $\kappa$ ensure stronger fairness in the sense of how large the majority of parties that \op{of-broadcast} some $\tx$ before $\tx'$
must be to ensure that $\tx$ will be \op{of-delivered} before $\tx'$ and in a fair order.

The protocol consists of three budling blocks: FIFO consistent broadcast channel, validated Byzantine consensus, and graph module. Since the FIFO consistent broadcast channel and validated Byzantine consensus will be described in Section~\ref{sec:implementation}, we will focus on the graph-building phase in this section. Concretely, we will describe how the protocol builds a graph and determines the order of transactions using a toy example.

\subsection{Broadcast and consensus}
\begin{figure}[!ht]
  \begin{center}
    \includegraphics[width=1\linewidth]{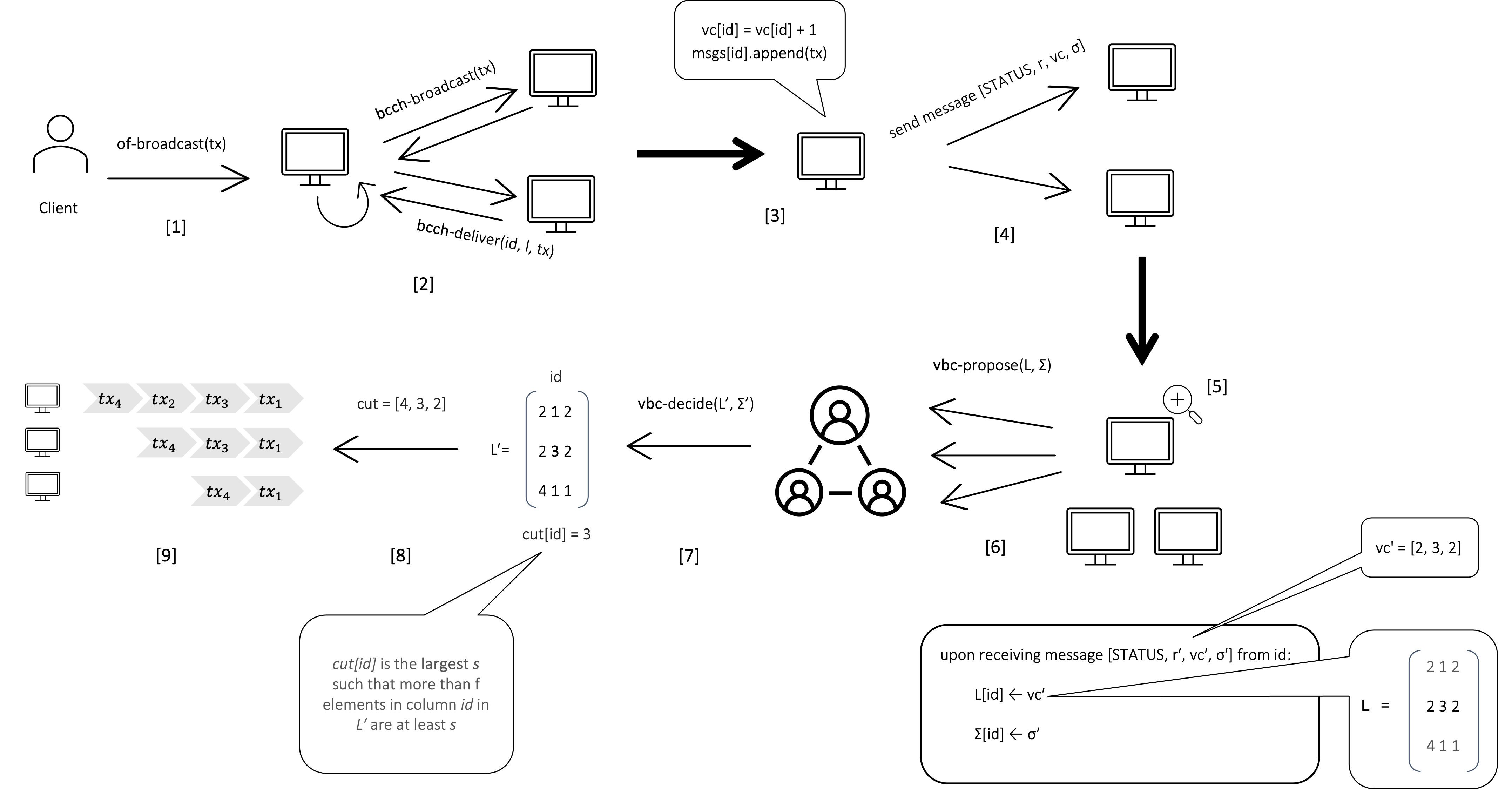}
  \end{center}
  \caption{Execution of the protocol shows the first (consensus) phase of the protocol.}
  \label{fig:consensus}
\end{figure}

The quick order fairness protocol proceeds in rounds and concurrently uses a FIFO consistent broadcast channel (bcch) to deliver transactions~\cite{DBLP:conf/fc/CachinMSZ22}. Figure~\ref{fig:consensus} depicts an example of the first phase of the protocol. The system consists of three correct parties $p_i, p_j, p_k$, and $\kappa$ is zero. To keep the example simple, we do not include Byzantine parties and consider the first round. 

The protocol starts once the client submits a transaction $\tx_1$ (1). Over time, the client will submit three more transactions: $tx_2, tx_3$, and $tx_4$ (not necessary in this order). The submitted transaction is \emph{bcch-broadcasted} (2) to other parties. Every party keeps a local vector clock \vc that counts the transactions that have been \emph{bcch-delivered} from each sending party. Every party also maintains an array of lists \msgs such that $\msgs[i]$ records all \emph{bcch-delivered} transactions from $p_i$. Upon \emph{bcch-delivering} $\tx$, every party increments \vc counter and appends the transaction to local array \msgs (3). The new round starts when sufficiently many new transactions are found in \msgs. In the next step, each party signs its \vc and sends it to all other parties as \str{status} message (4). All received vector clocks and signatures are stored in matrices $L$ and $\Sigma$ (5). A row of matrix $L$ maps to received \vc from the party $\var{id}$. Once $n-f$ \str{status} messages are collected, a party proposes $L$ and $\Sigma$ to the consensus module (6).

The protocol then runs a validated Byzantine consensus (vbc) protocol to agree on a matrix $L'$ and a list $\Sigma'$ of signatures that validate $L'$ (7). In this example, the protocol decides on matrix $L'$ and uses it to determine the cut (8). Each matrix column calculates the largest value such that more than $f$ elements in the column are at least that value. The cut is then defined as the vector of these values corresponding to each party. The cut is $[4, 3, 2]$ in this example. The cut determines an entry in \msgs array up to which transactions are considered for creating the fair order in the round (9). A party may be missing some transactions in the cut. In that case, the protocol will ask other parties to send the missing transactions. Once the message exchange is completed, the protocol proceeds to the next phase.

\subsection{Building a graph}

The next phase is building a graph. Figure~\ref{fig:execution0} shows steps of building a directed dependency graph representing a fair transaction order. Previous execution steps produced the cut [4, 3, 2]. This means that in the current round (10), party $p_i$ observes $\tx_4 \prec \tx_2 \prec \tx_3 \prec \tx_1$, party $p_j$ observes $\tx_4 \prec \tx_3 \prec \tx_1$ and $p_k$ observes $\tx_4 \prec \tx_1$. The first step is to create vertices of the graph by selecting all unique transactions within the cut that have not yet been delivered (11). In this example, this will produce four vertices. The next step is constructing matrix $M$ (12), i.e., calculating for every party how many times transaction $\tx$ is delivered before $\tx'$, within the cut. The next step adds edges (13) to the graph by checking the following condition:
\begin{equation}
  \op{max}\{M[\tx][\tx'], n-f-M[\tx'][\tx] \} > M[\tx'][\tx]-f+\kappa
\end{equation}
where $M[\tx][\tx']$ is the number how many times is $\tx$ delivered before $\tx'$, $f$ is the number of faulty parties, $n$ is the total number of parties and $\kappa$ is fairness parameter. Note that this condition check is done for each pair of vertices so that edges might be added in both directions. To avoid the problem of Condorcet cycles, the next step tries to collapse the graph (14). In this example, there is a path from every vertex to another, so the whole graph is collapsed into a single vertex. Starting from the vertex with zero incoming edges, the protocol extracts transactions from the vertex and checks \emph{stable} condition, i.e., if every transaction appears more or equal to $\frac{n+f-\kappa}{2}$ times within the cut. In this example, transaction $\tx_2$ appears only once in the cut but should appear at least twice to be stable (15). Therefore, the protocol cannot deliver anything and moves to the next round (16).

\begin{figure}[!ht]
  \begin{center}
    \includegraphics[width=1\linewidth]{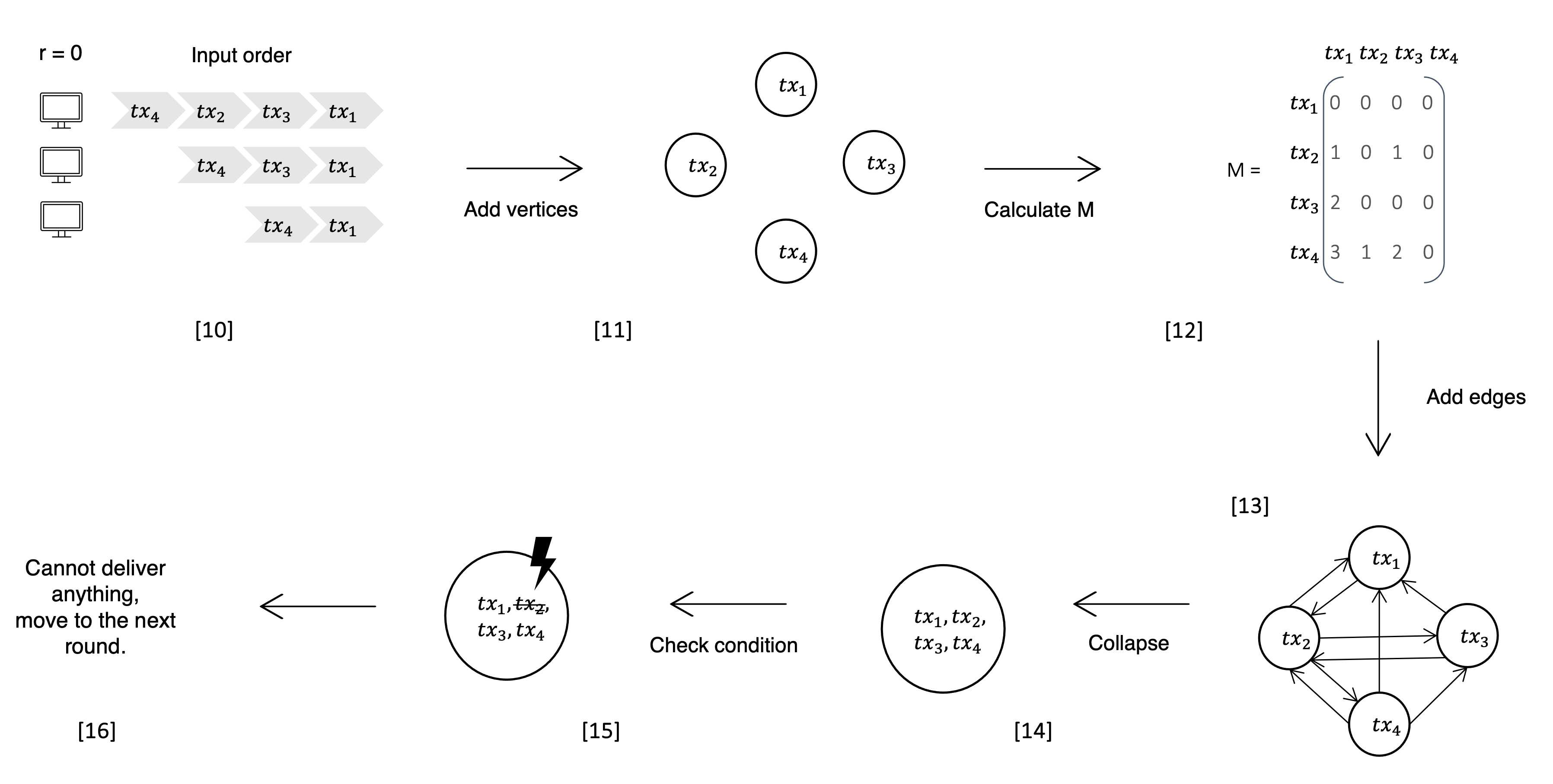}
  \end{center}
  \caption{Example execution of building a graph starting from the first round. }
  \label{fig:execution0}
\end{figure}

Figure~\ref{fig:execution1} shows a continuation of the execution shown in Figure~\ref{fig:execution0}. Meanwhile, more transactions arrived at parties (17), extending the cut to [4, 4, 4]. As in the previous figure, we add vertices (18), calculate again matrix M (19), and add edges (20) to the graph following the same steps. The difference from the last round is that this time, we see fewer edges in the graph because the protocol has more information about the order, making it more confident about transaction ordering. After the collapsing stage (21), we create two vertices ($\tx_4$) and ($\tx_1, \tx_2, \tx_3$). The protocol starts from vertex ($\tx_4$) (with zero incoming edges) and checks \emph{stable} condition (22). That is more than enough since the transaction $\tx_4$ appears four times in the cut. It is \op{of-delivered} first. Then, we remove vertex $\tx_4$ from the graph (23), and the protocol chooses the next vertex ($\tx_1, \tx_2, \tx_3$). Again, because the next vertex has multiple transactions inside, the protocol checks if each satisfies \emph{stable} condition. We deliver all three transactions together this time since they fulfill the condition (24). We then remove the corresponding vertex, leaving nothing else to deliver.

\begin{figure}[!ht] 
  \begin{center}
    \includegraphics[width=1\linewidth]{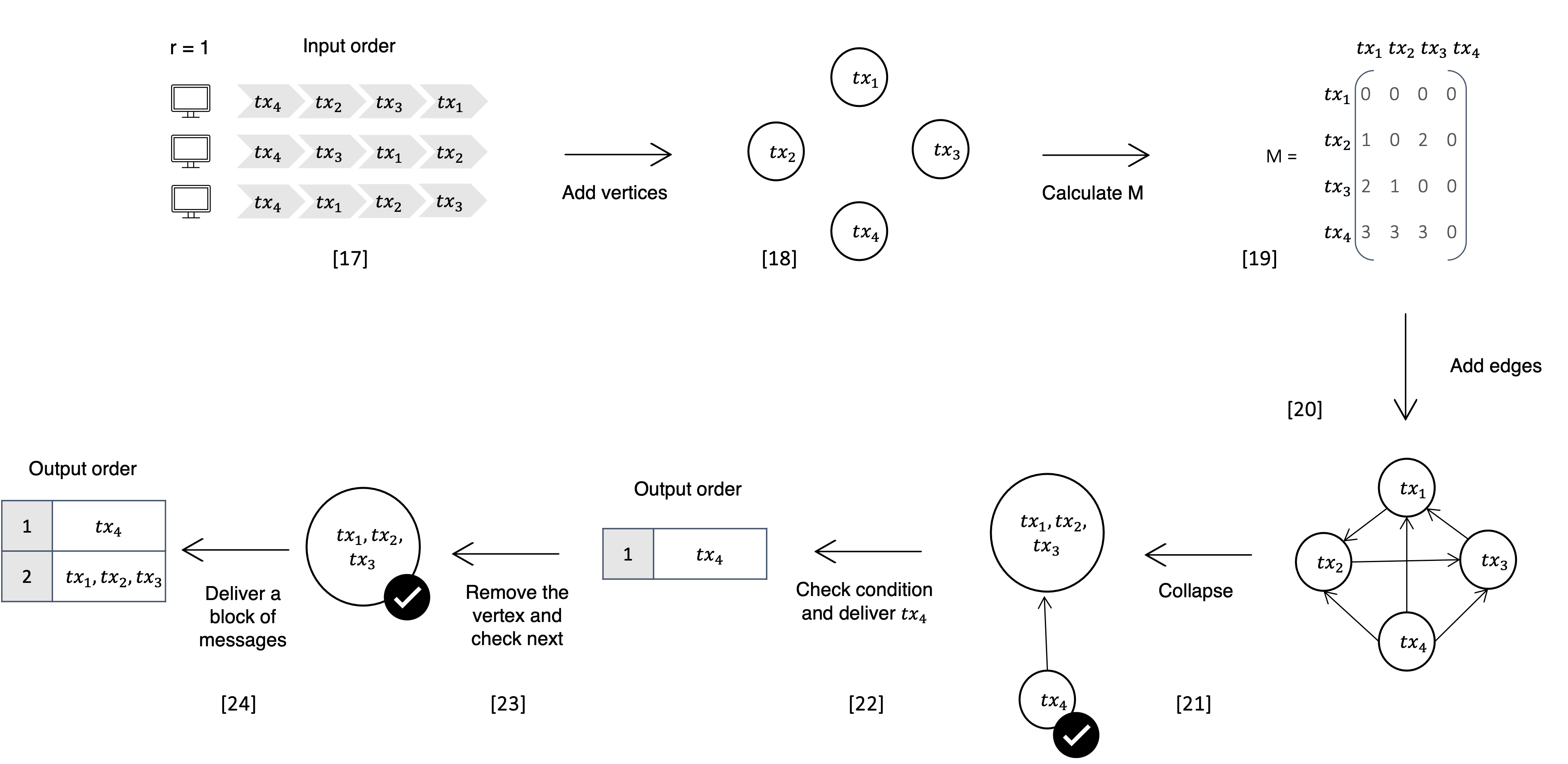}
  \end{center}
  \caption{Continuation of the execution from Figure~\ref{fig:execution0}.}
  \label{fig:execution1}
\end{figure}

\section{Implementation}
\label{sec:implementation}
Implementing the quick order-fair protocol builds on top of three modules: a Byzantine consistent broadcast module, a validated Byzantine consensus module, and a graph module. In the following, we describe the implementation of these modules. The code is written in Go version 1.15.7.

\subsection{Byzantine Consistent Broadcast Channel}
We modularly implement the Byzantine consistent broadcast channel (bcch) abstraction. For every sender, bcch invokes a sequence of broadcast primitives (bcb) so that only one is active at any moment. BCB relies on authenticated perfect links (al) that communicate with Transmission Control Protocol (TCP).

\begin{figure}[!ht]
  \centering
  \includegraphics[width=0.6\textwidth]{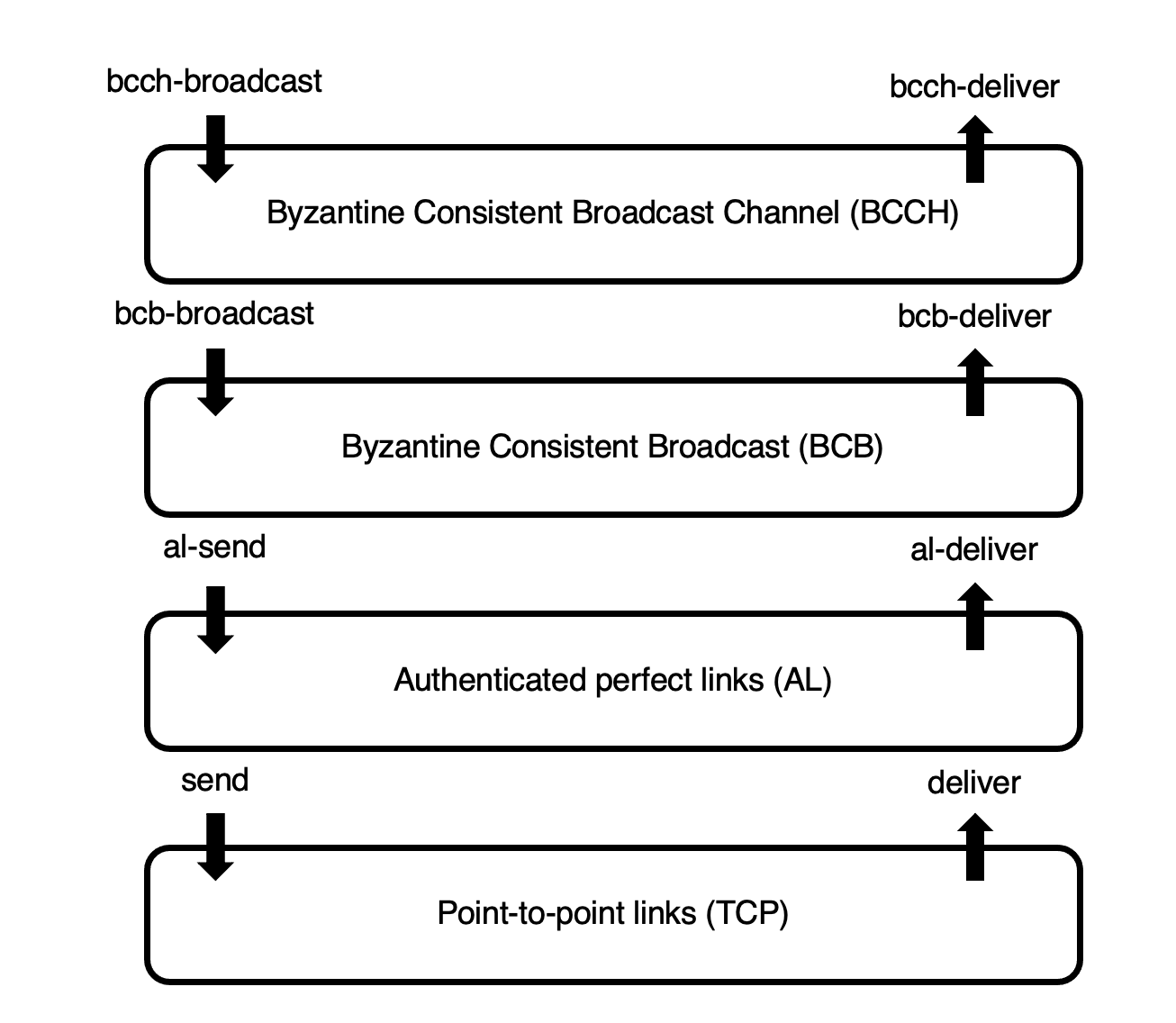}
  \caption{A stack of modules for implementing Byzantine consistent broadcast channel in Quick Order Fairness protocol.}
  \label{fig:bcch}
\end{figure}

\subsubsection{Implementation details}

The Byzantine consistent broadcast channel implementation follows Algorithm 3.19 of Cachin \etal~\cite{DBLP:books/daglib/0025983}. Initially, each party creates an instance of bcch, automatically creating an instance of Byzantine consistent broadcast (\emph{bcb}). Then protocol waits for \op{of-broadcast($\tx$)} event to happen. Every time this event is triggered, \op{bcch-broadcasts} $\tx$ using underlying \op{bcb-broadcast($\tx$)} primitive. This primitive implements Signed Echo Broadcast presented in Algorithm 3.17 \cite{DBLP:books/daglib/0025983}. This protocol uses an authenticated perfect links abstraction and a cryptographic digital signature scheme. Authenticated perfect links (\emph{al}) primitive implements Authenticate and Filter shown in Algorithm 2.4 \cite{DBLP:books/daglib/0025983}. It uses a Hash-based message authentication\footnote{HMAC is a cryptographic technique with a hash function and a secret key.} (HMAC) over a TCP network communication. Specifically, in our implementation, we implement HMAC\textunderscore256 in the file \emph{hmac.go}, package \emph{hotstuff-impl/crypto}.

\subsection{Validated Byzantine Consensus}
This module implements validated Byzantine consensus (vbc) introduced by Cachin \etal~\cite{DBLP:conf/crypto/CachinKPS01} using the HotStuff~\cite{DBLP:conf/podc/YinMRGA19} implementation in the project \emph{bamboo}\footnote{\url{https://github.com/gitferry/bamboo}}. Observe that HotStuff does not provide validation. We must modify the implementation to cope with the \emph{external validity} property. Moreover, HotStuff is implemented as an atomic broadcast instance: the output for every correct party is a sequence of ordered transactions. To make this into a consensus protocol, we agree on considering the first message output by a correct party proposed by the leader for round $r$. 

\begin{figure}[!ht]
  \centering
  \includegraphics[width=0.6\textwidth]{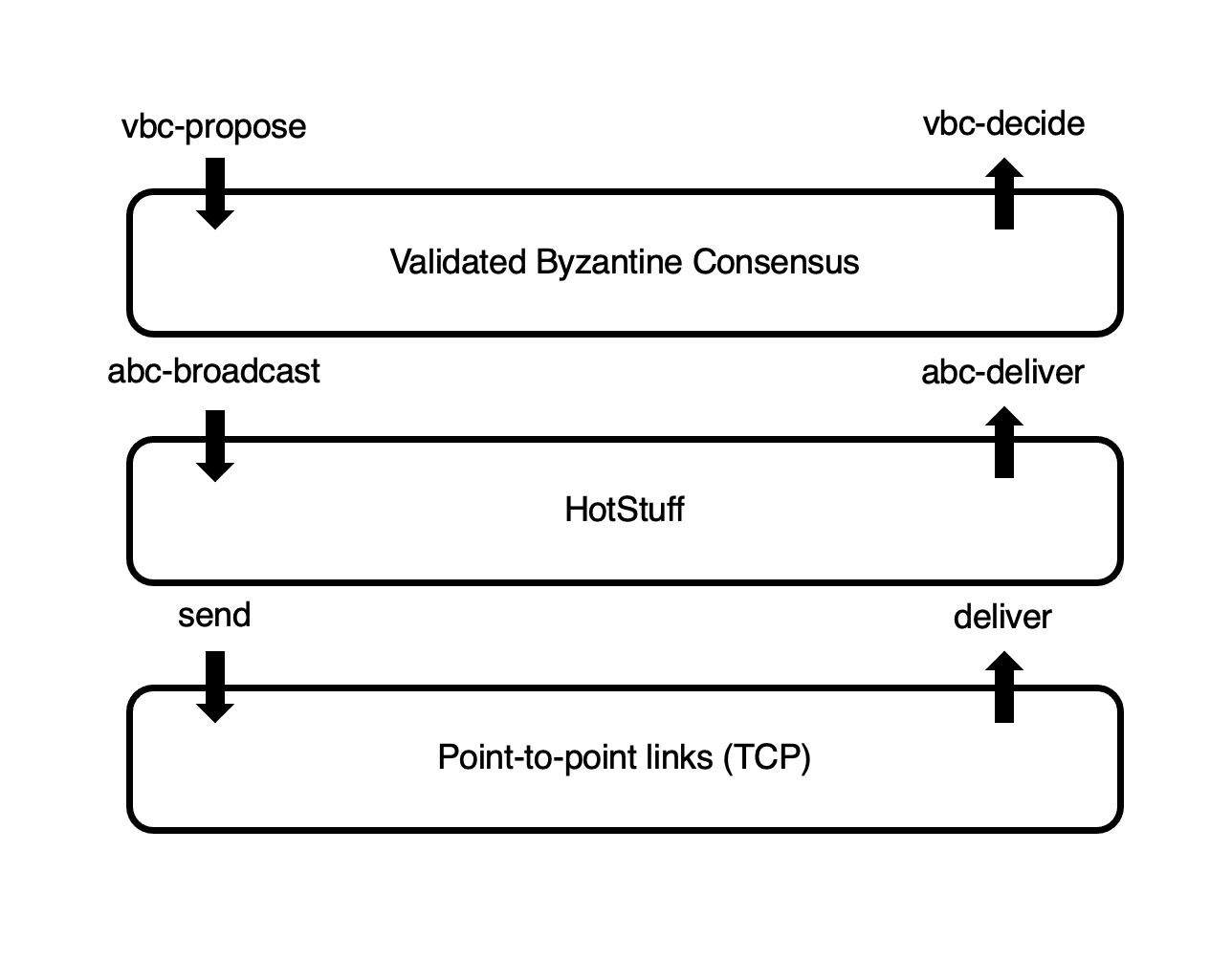}
  \caption{A stack of modules for implementing validated Byzantine consensus in Quick Order Fairness protocol.}
  \label{fig:vbc}
\end{figure}

Figure~\ref*{fig:vbc} shows a high-level architecture for implementing validated Byzantine consensus. A client sends a \emph{of-broadcast} transaction to the Quick Order Fairness module. This module communicates with the validated Byzantine consensus module through a \emph{vbc-propose}. 

A validated Byzantine consensus protocol is activated by a \emph{vbc-propose} message carrying a value~$v$ with a proof $\pi$ that validates $v$, i.e., $\pi$ should satisfy a predicate $P$ for $v$. A correct party only decides for values validated by a proof $\pi$. In our quick order-fair protocol, a correct party proposes for consensus a matrix $L$ of vector clocks (counting how many transactions were \op{bcch-delivered} from each party) together with a list $\Sigma$ containing the signatures of the parties on the vector clocks, which represents the proof $\pi$ for the validation. We define predicate $P$ as true whenever $\Sigma$ correctly verifies $L$ for round $r$. In particular, 

$$P_r(L,\Sigma) = \true \text{ iff } \forall \var{vc}' \in L,~\forall \sigma' \in \Sigma:\op{verify}(j,\var{vc}',\sigma').$$

The HotStuff protocol implementation verifies that $P_r(L,\Sigma)$ holds whenever the leader for round $r$ broadcasts a proposal block containing a list of (possibly different) matrices $L$ and whenever a correct replica delivers the block proposed by the leader; proceed only if $P_r(L,\Sigma) = \true$ and halt otherwise. HotStuff uses underlying point-to-point links for internal communication. The first matrix in the delivered block is decided and used to determine the $\var{cut}$ for round $r$. More precisely, the validation party occurs in the validated Byzantine consensus module, which \emph{vbc-decides} on a matrix $L'$ with a proof $\pi$ that validates $L'$.

\subsubsection{Implementation details} An abstract vbc module (Module~\ref{alg:vbcmodule}) has the following events: \emph{vbc-propose($v$)} and \emph{vbc-decide($v$)}. A correct party first proposes a value and then must wait for a decision before it proposes a new value.

\begin{module*}
  \vbox{
  \footnotesize
  \begin{numbertabbing}
    xxxx\=xxxx\=xxxx\=xxxx\=xxxx\=xxxx\=MMMMMMMMMMMMMMMMMMM\=\kill
    \textbf{Events:}\\
    \> \textbf{Request:} \eventmodp{vbc}{propose}{$v$}: Proposes value $v$.\\
    \> \textbf{Indication:} \eventmodp{vbc}{decide}{$v$}: Outputs a decided value $v$ of consensus.\\
    \textbf{Properties:}\\
    \> \emph{Agreement}: If some correct party decides $v$, then any correct party that terminates decides $v$.\\
    \> \emph{External Validity}: Any correct party that terminates decides $v$ such that proof $\pi$ satisfy a predicate $P$ for $v$.\\
    \> \emph{Integrity}: If all parties follow the protocol, and if some party decides $v$, then some party proposes $v$.\\
    \> \emph{Termination}: Every correct party eventually decides some value.\\
\end{numbertabbing}
}
\caption{Validated Byzantine consensus module.}
\label{alg:vbcmodule}
\end{module*}

 Module~\ref{alg:abcmodule} depicts the specification of the atomic broadcast abstraction. It has two events: \op{abc-broadcast}($v$) that broadcasts a value $v$ and \op{abc-deliver}($v$) that delivers a decided value $v$.

\begin{module*}
  \vbox{
  \footnotesize
  \begin{numbertabbing}
    xxxx\=xxxx\=xxxx\=xxxx\=xxxx\=xxxx\=MMMMMMMMMMMMMMMMMMM\=\kill
    \textbf{Events:}\\
    \> \textbf{Request:} \eventmodp{abc}{broadcast}{$v$}: Broadcasts a value $v$ to all parties.\\
    \> \textbf{Indication:} \eventmodp{abc}{deliver}{$v$}: Delivers a decided value $v$.\\
    \textbf{Properties:}\\
    \> \emph{Validity}: If a correct party $p$ broadcasts a value $v$, then $p$ eventually
    delivers $v$.\\
    \> \emph{No duplication}: No value is delivered more than once.\\
    \> \emph{No creation}: If a party delivers a value $v$ with sender $s$, then $v$ was
    previously broadcast by party $s$.\\
    \> \emph{Agreement}: If a value $v$ is delivered by some correct party,\\
    \>then $v$ is eventually delivered by every correct party.\\
    \> \emph{Total order}: Let $v$ and $v'$ be any two values and suppose $p$ and $q$ are
    any two correct parties that deliver $v$ and $v'$. \\
    \> If $p$ delivers $v$ before $v'$, then $q$ delivers $v$ before $v'$.\\
\end{numbertabbing}
}
\caption{Atomic broadcast (HotStuff)}
\label{alg:abcmodule}
\end{module*}

\begin{table}[!ht]
  \parbox{.45\linewidth}{
  \centering
  \begin{tabular}{ |l|l| } 
   \hline
   \textbf{Attribute} & \textbf{Description}  \\ 
   \hline
   vbc & Message tag.  \\ 
   $v$ & Proposed value.  \\
   \rp & Proposing round.  \\ 
   \hline
  \end{tabular}
  \caption{Structure of a proposed transaction.}
  \label{tbl:transaction struct}
  }
  \hfill
  \parbox{.50\linewidth}{
    \centering
  \begin{tabular}{ |l|l| } 
   \hline
   \textbf{Attribute} & \textbf{Description}  \\ 
   \hline 
   \cview & Round in which is the vote created. \\ 
   \var{voter} & Party that created the vote.\\
   \bid & Identifier of a voted block.  \\
   $\sigma$ & Signature of the vote.  \\
   \hline
  \end{tabular}
  \caption{Structure of a vote.}
  \label{tbl:vote struct}
  }
\end{table}

Algorithm~\ref{alg:alg1} shows an abstract implementation of vbc using atomic broadcast (ABC) running on $p_i$. A flag \inround is, by default, set to false and will change only when a consensus round starts. Variables \rp and \rd count how many times a value is proposed, respectively, and delivered. A new vbc consensus round starts when \op{vbc-propose}($v$) is triggered only when \inround is false. Then, counter \rp increases by one, \inround is set to true, and the transaction is broadcasted. Transaction structure is given in Table~\ref{tbl:transaction struct}. Upon party $p_i$ receiving transaction $w$ (containing value $v$) from another party, it checks if the round of received transaction is the same as its \rp and if $\rp > \rd$. If yes, then \rd is increased by one, \inround set to false, and value $v$ is \op{vbc-decided}. Otherwise, the proposal value $v$ is discarded.

\begin{algo*}[!ht]
  \vbox{
  \footnotesize
  \begin{numbertabbing}
    xxxx\=xxxx\=xxxx\=xxxx\=xxxx\=xxxx\=MMMMMMMMMMMMMMMMMMM\=\kill
    \textbf{State:}\\
    \> $\inround \gets \false$ \` // flag if we can start consensus \label{}\\
    \> $\rp \gets 0$  \` // counter for how many times proposed (same as QOF round) \label{}\\
    \> $\rd \gets 0$ \` // counter for how many times delivered \label{}\\
    \\
    \textbf{upon} $\op{\vbc-propose}(v)$ such that not $\inround$ \textbf{do} \label{}\\
    \> $\rp \gets \rp + 1$ \` //in QOF this happens after building the graph\label{}\\
    \> $\inround \gets \true$ \label{}\\
    \> $\op{abc-broadcast}([\str{vbc}, \rp, v])$ \` // abc-broadcast proposal of $p_i$ for vbc instance $\rp$ \label{}\\
    \\
    \textbf{upon} $\op{abc-deliver}(w)$ such that $w = [\str{vbc}, r, v]$ \textbf{do} \label{}\\
    \> \textbf{if} $r = \rp \land \rp > \rd$ \textbf{then}
    \` // first \op{abc-delivered} proposal for this round \label{}\\
    \>\> $\rd \gets \rd + 1$ \label{}\\
    \>\> $\inround \gets \false$ \label{}\\
    \>\> $\op{\vbc-decide}(v)$ \label{}\\
    \> \textbf{else}\label{}\\
    \>\> // discard proposal value~$v$ \label{}
  \end{numbertabbing}
  }
  \caption{Abstract implementation of vbc using ABC (code for $p_i$).}
  \label{alg:alg1}
\end{algo*}

Algorithm~\ref{alg:alg2} is a concrete implementation of vbc in QOF within HotStuff of the bamboo library. As in the previous algorithm, \rp and \rd keep consensus running correctly. Additionally, \cview keeps track of HotStuff round, and matrix \votes stores votes for a specific \bid, given by \var{voter}. Upon \op{vbc-propose} with value $v$, the algorithm starts a new round of vbc consensus by increasing \rp by one and creates transaction $t$, which is a tuple of \str{vbc} tag, \rp and $v$. The created transaction is added to the local mempool of the bamboo library of a party $p_i$ using the \op{mempool.addNew}($t$) function. 

When party $p_i$ becomes leader, it increases \cview by one and creates a block. Block is created from payload generated by the function \op{mempool.some}(). This function will get up to 20 transactions from the mempool and put them in \payload. Then, the function \op{makeBlock} takes \payload and \cview to build a block. The block is then sent to all parties in the message \str{block}. 

When party $p_i$ receives a block from some other party, it uses it to create a vote. First, using cryptographic function \op{sign}, party $p_i$ signs \bid and produces signature $\sigma$, which is included in \vote. The vote structure is given in Table~\ref{tbl:vote struct}. Then, the created vote is sent inside \str{vote} message to the next leader. 

When a leader receives the vote, it first verifies it, using the \op{verify} function,  and only then it adds the vote into matrix \votes. If more than $\frac{2}{3}$ of a total number of parties voted for the same \bid, a quorum is reached, and a quorum certificate (\qc) is generated. At the same time, \cview is updated. The block can be committed if the \qc view is bigger or equal to three. As we know, HotStuff takes three rounds to commit a block, so the previous condition comes from this fact. Before committing the block, the last check is to check if the view of the grandparent's and parents' blocks is correct. Finally, the block is committed by calling the function \op{commitBlock} that takes the grandparent block and current view as arguments. This function will append a committed block to the channel called \cblocks. 

Every time a new block is committed, the algorithm takes the last block from \cblocks and extracts the payload in transaction $t$. If an extracted tag is \str{vbc}, the proposal round of $t$ is the same as the current proposal round, and $\rp > \rd$ then \rd is increased by one, and the value stored in $t$ is \op{vbc-decided}. Otherwise, the value of the proposed transaction, respectively, is discarded.

\begin{algo*}[!ht]
  \vbox{
  \footnotesize
  \begin{numbertabbing}
    xxxx\=xxxx\=xxxx\=xxxx\=xxxx\=xxxx\=MMMMMMMMMMMMMMMMMMM\=\kill
    \textbf{State:} \\
    \> $\rp \gets 0$: counter of how many times value is proposed (same as QOF round) \label{}\\
    \> $\rd \gets 0$: counter of how many times value is decided \label{}\\
    \> $\cview \gets 0$: counter of HotStuff protocol rounds \label{}\\
    \> $\votes \gets [0]^{n\times n}$: matrix of votes for each \bid and \var{voter} \` // votes[vote.BlockID][vote.Voter]\label{}\\
    \\
    \textbf{upon} $\op{\vbc-propose}(v)$ \textbf{do} \label{}\\
    \> $\rp \gets \rp + 1$ \` //in QOF this happens after building the graph \label{}\\
    \> // instead of \op{abc-broadcast}, add the proposal of $p_i$ for the vbc instance as a new transaction to the mempool \label{}\\
    \> $t \gets (\str{vbc}, \rp, v)$ \label{}\\
    \> $\op{mempool.addNew}(t)$ \label{}\\
    \\
    \textbf{upon} becoming leader \textbf{do} \label{}\\
    \> $\cview \gets \cview + 1$ \label{}\\
    \> $\payload \gets \op{mempool.some}()$ \`// get up to 20 transactions \label{}\\
    \> $\block \gets \op{makeBlock}(\payload, \cview)$ \label{}\\
    \> send message \([\str{block}, \block] \) to all \(p \in \CP\) \`// broadcasts block; vbc.go L348 \label{}\\
    \\
    \textbf{upon} receiving message \([\str{block}, \block] \) from $p_j$ \textbf{do} \label{}\\
    \> $\sigma \gets \op{sign}(i, \bid)$\label{}\\
    \> $\vote \gets \op{makeVote}(\cview, i, \bid, \sigma)$ \`// hotstuff.go L105 \label{}\\
    \> send message   \([\str{vote}, \vote]\) to next view \`// to next leader; hotstuff.go L105 \label{}\\    \\  
    \textbf{upon} receiving message \([\str{vote}, \vote]\) from $p_j$ \textbf{such that} $\op{verify}(j, \vote.\bid, \vote.\sigma)$ \textbf{do} \`// hotstuff.go L123 \label{}\\
    \> $\votes[\vote.\bid][\vote.\var{voter}] \gets \vote$\label{}\\
    \> \textbf{if} $\#(\votes[\vote.\bid]) > \#(\CP) \cdot \frac{2}{3}  $ \textbf{then} \`// quorum.go L66\label{}\\
    \>\> $\qc \gets (\vote.\cview, \vote.\bid)$\label{}\\
    \>\> $\cview \gets \qc.\view + 1$  \label{}\\
    \>\> \textbf{if} $\qc.\view \geq 3$ \textbf{then}\label{}\\
    \>\>\> $\pb \gets \op{getParentBlock}(\qc.\bid)$  \`// hotstuff.go L287\label{}\\
    \>\>\> $\gb \gets \op{getParentBlock}(\pb.\bid)$  \`// hotstuff.go L291\label{}\\
    \>\>\> \textbf{if} $\gb.\view+1 = \pb.\view$ \textbf{and} $\pb.\view+1 = \qc.\view$ \textbf{then}\label{}\\
    \>\>\>\> // instead of \op{abc-deliver}, a block is appended to \cblocks \label{}\\
    \>\>\>\> $\cblocks \gets \op{commitBlock}(\gb, \cview)$ \label{}\\
    \\
    \textbf{upon} \cblocks is updated \textbf{do} \label{}\\
    \> $\cblock \gets \cblocks[0]$ \`// get latest committed block \label{}\\
    \> $t \gets \cblock.payload[0]$ \label{}\\
    \> $(\var{tag}, \var{round}, \var{val}) \gets t$ \label{}\\
    \> \textbf{if} $\var{tag} = \str{vbc} \land \var{round} = \rp \land \rp > \rd$ \textbf{then} \label{}\\
    \>\> $\rd \gets \rd + 1$ \label{}\\
    \>\> $\op{\vbc-decide}(\var{val})$\`// vbc.go L299 \label{}\\
    \> \textbf{else} \label{}\\
    \>\> // discard proposed value \label{}

  \end{numbertabbing}
  }
  \caption{Concrete implementation of vbc in QOF within HotStuff of bamboo (code for $p_i$).}
  \label{alg:alg2}
\end{algo*}

\subsection{Graph Building}
The last phase of the quick order fairness protocol is building a graph that reflects the fair order of transactions. The graph is implemented as a directed acyclic graph (DAG), where every vertex represents a transaction (delivered by bcch), and an edge represents a dependency between two transactions. Package \emph{graph} holds the implementation of the graph structure and utility functions.
 
\subsubsection{Implementation details}

\begin{table}[!ht]
  \begin{center}
  \begin{tabular}{|l|l|} 
   \hline
   \textbf{Function} & \textbf{Description}\\
   \hline
    NewDirectedGraph & Creates a new directed graph.\\
    AddVertex & Adds a vertex to a graph.\\
    RemoveVertex & Removes a vertex from a graph.\\
    AddEdge & Adds an edge between two vertices. \\
    CollapseGraph &  Collapses a graph into a single vertex.\\
    SCC & Implements Strongly Connected Components. \\
    DFS & Implements Depth First Search. \\
    Transpose & Transposes a graph. \\
    Visit & Loops through the graph using DFS and outputs SCC. \\
    Indegree & Calculates the indegree of a vertex. \\

   \hline
  \end{tabular}
  \caption{\label{tbl:graphfunctions}List of implemented graph utility functions.}
  \end{center}
\end{table}

Algorithm~\ref{alg:graph} depicts the structure of the implemented graph. The graph is defined as a map of vertices. The Vertex structure represents each vertex. The relations between vertices represent the graph's edges,i.e., an edge has no explicit structure. Each vertex keeps track of outbound edges to other vertices. 

Implemented functions in the graph module are given in Table~\ref{tbl:graphfunctions}. These functions create a graph, add and remove vertices, add edges, collapse a graph, calculate strongly connected components, and implement other helper functions. This paper will focus on the implementation of collapsing a graph functionality, given in Algorithm~\ref{alg:graph}, since it is the most complex function in the graph module.

After a party constructs vertices and edges of the graph $G$, it will call function \op{CollapseGraph} that will try to collapse the graph $G$ into a single vertex (L\ref{collapse}). The function \op{CollapseGraph} first checks if the graph has less or equal to one vertex. If yes, it returns the same graph since there is nothing to collapse. Otherwise, it creates a new graph $H$ and calculates strongly connected components (SCC) of $G$ (L\ref{SCC}). Then, it loops through all components and adds them as vertices to the new graph $H$. Finally, it returns the new graph $H$.

The function \op{SCC}($G$) finds strongly connected components, i.e., in a directed graph, it checks if there is a directed path from every vertex in the component to every other vertex in the same component (L\ref{SCC}). It loops through all vertices in $G$ and checks if a vertex has been visited. If this is not the case, it calls \op{DFS} function (L\ref{DFS}) that will traverse the graph and mark all visited vertices. Then, it transposes the graph $G$ and loops through all vertices in the stack. For each vertex from the stack, it checks if it has already been visited. If not, it calls \op{Visit} function that will loop through the graph using DFS and fill \cmp list. Finally, it returns the \cmp.

Function \op{DFS}($\visited, \stack, \node$) implements Depth First Search (DFS) algorithm (L\ref{DFS}). It checks if a given \node is already visited. If not, it marks it as visited and loops through all neighbors of the \node. For each neighbor, it calls the recursively \op{DFS} function. Finally, it appends the vertex to the stack. We choose to implement DFS since it is a simple algorithm with efficient time complexity.

\begin{algo*}[!ht]
  \vbox{
  \footnotesize
  \begin{numbertabbing}
    xxxx\=xxxx\=xxxx\=xxxx\=xxxx\=xxxx\=MMMMMMMMMMMMMMMMMMM\=\kill
    \textbf{Type:} \label{}\\
    \> Graph( \label{}\\
    \>\> HashMap[string $\rightarrow$ Vertex] \ver) \`// Collection of vertices in the graph. \label{}\\
    \> Vertex( \label{}\\
    \>\> string \var{key}, \`// Unique identifier of a vertex. \label{}\\
    \>\> bcchMessage \var{data}, \`// Transaction delivered by bcch. \label{}\\
    \>\> HashMap[string $\rightarrow$ Vertex] \ver) \`// Collection of connected vertices. \label{}\\
    \> bcchMessage( \label{}\\
    \>\> []byte \var{message}, \`// Transaction delivered by bcch. \label{}\\
    \>\> int \var{round}, \`// Round in which the message was delivered. \label{}\\
    \>\> int \var{fromProcess}, \`// Party that delivered the message. \label{}\\
    \>\> string \var{id}) \`// Unique identifier of a message. \label{}\\
    \\
    \textbf{upon} $\op{CollapseGraph}(G: \text{Graph})$ \textbf{do} 
    \label{collapse}\\
    \> \textbf{if} $\#(G.\ver) \leq 1$ \textbf{then} \label{}\\
    \>\> \textbf{return} $G$ \label{}\\
    \> \textbf{else} \label{}\\
    \>\> $H \gets \op{NewDirectedGraph}()$ 
    \label{}\\
    \>\> $\cmp \gets \op{SCC}(G)$ \label{}\\
    \>\> \textbf{for} $c \in \cmp$ \textbf{do} \label{}\\
    \>\>\> $H.\op{AddVertex}(c)$ 
    \label{}\\
    \>\> \textbf{return} $H$ \label{}\\
    \\
    \textbf{function} $\op{SCC}(G: \text{Graph})$ 
    \label{SCC}\\
    \> $\ver \gets G.\ver$ \label{}\\
    \> $\visited, \cmp, \stack \gets \emptyhashmap$ \label{}\\
    \> \textbf{for} $i \in \{0,\dots, \#(\ver)\}$ \textbf{do}\label{}\\
    \>\> $\node \gets \ver[i]$ \label{}\\
    \>\> \textbf{if} $\visited[node] = \false$ \textbf{then}\label{}\\
    \>\>\> \op{DFS}(\visited, \stack, \node) \label{}\\
    \> $\transposed \gets G.\op{Transpose}()$ 
    \label{}\\
    \>$\visited \gets \emptyhashmap$ \label{}\\
    \>\textbf{for} $\#(\stack) \neq 0$ \textbf{do} \label{}\\
    \>\> $v \gets \stack[\#(\stack)-1]$\label{} \`// removes and returns the last element\\
    \>\> \textbf{if} $\visited[v] = \false$ \textbf{then} \label{}\\
    \>\>\> \transposed.\op{Visit}(\visited, $v$, \cmp) 
    \label{}\\
    \>\textbf{return} \cmp \label{}\\
    \\
    \textbf{function} \op{DFS}(\visited, \stack, \node) 
    \label{DFS}\\
    \> \textbf{if} $\visited[\node] = \false$ \textbf{then} \label{}\\
    \>\>$\visited[\node] \gets \true$ \label{}\\
    \>\> \textbf{for} $\neighbour \in \node.\ver$  \textbf{do}\label{}\\
    \>\>\> \op{DFS}(\visited, \stack, \neighbour)\label{}\\
    \>\> $\stack.\op{append}(\node)$\label{}
  \end{numbertabbing}
  }
  \caption{Abstract implementation of collapsing a graph.}
  \label{alg:graph}
\end{algo*}

\section{Integration}
\label{sec:integration}
This section discusses how to deploy QOF in a blockchain network. One approach is implementing it as a separate service through which clients submit transactions. Alternatively, the network's validators can integrate it directly with the consensus protocol.

\begin{figure}[!ht]
  \centering
  \includegraphics[width=1.0\textwidth]{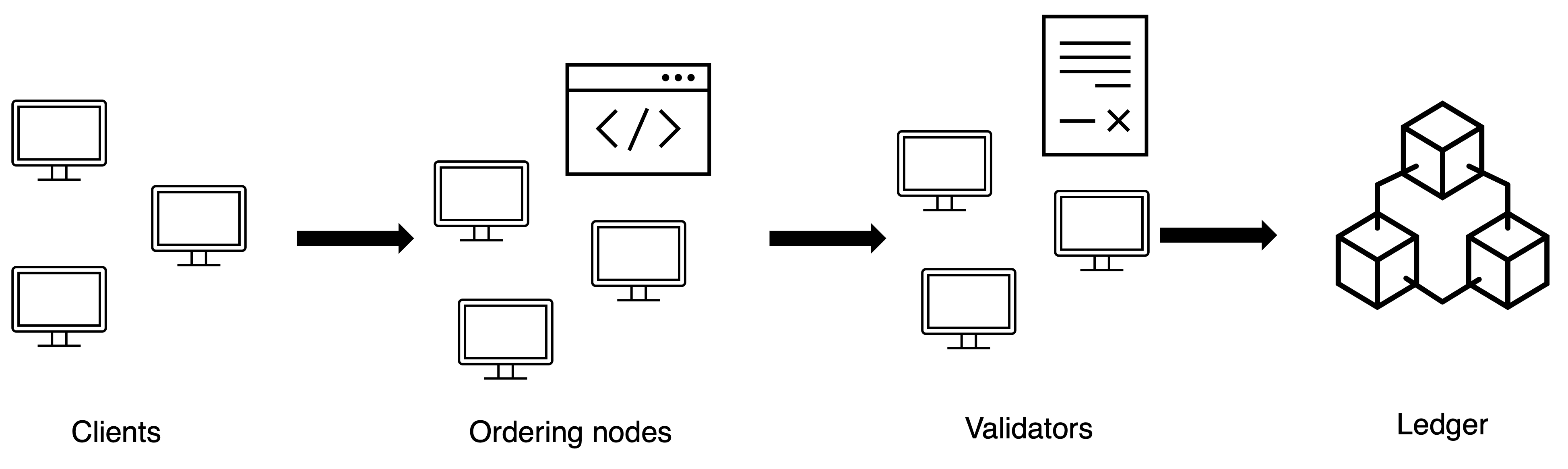}
  \caption{Clients submit transactions to ordering nodes that use the QOF algorithm as a separate service. The algorithm outputs transactions in fair order and sends them to the validators who run a smart contract. Smart contract executes transactions on the ledger in a fair order.}
  \label{fig:int1}
\end{figure}

\paragraph{A separate service.}
In one approach, clients first send transactions to specific \emph{ordering nodes} responsible for imposing order fairness (Figure~\ref{fig:int1}). In this scenario, the QOF protocol provides a separate ordering mechanism, like those implemented by layer-two networks. In contrast to many layer-two solutions, the ordering nodes implement proper distributed consensus. The ordering nodes output a sequence of transactions that respects differential order fairness. This information is signed and sent to validators who run the smart contract that executes transactions on the ledger in the given order. With this approach, the ledger protocol is agnostic to the fair ordering process, and a contract may opt to consume only transactions in a fair order. In this case, the smart contract should verify the digital signatures of the ordering nodes before executing transactions. Some pseudocode for a smart contract of this kind is shown in Algorithm~\ref{alg:smartcontract}. 

This approach can readily be integrated with many layer-two networks that have recently become popular~\cite{l2beat}. These networks increase scalability and efficiency by moving transaction execution off the main blockchain.  Running off-chain, they enable faster and more cost-effective transactions while retaining the security benefits of the underlying blockchain by pushing only the effects of these transactions onto the main network.  Today, layer-two systems typically use a centralized sequencer that gathers client transactions and organizes them. Subsequently, transactions are executed through the rollup mechanism, and the resulting updated state is recorded on layer one, the main blockchain, through a rollup contract deployed there. As most employ just one sequencer, this poses a single point of failure and restricts interoperability with other, distinct layer-two systems. It is readily possible to distribute the function of the sequencer across multiple nodes, even to serve multiple rollup protocols from the same distributed layer-two sequencer. Such an approach was recently introduced by \emph{Espresso}~\cite{espresso}.

\begin{algo*}[!ht]
  \vbox{
  \footnotesize
  \begin{numbertabbing}
    xxxx\=xxxx\=xxxx\=xxxx\=xxxx\=xxxx\=MMMMMMMMMMMMMMMMMMM\=\kill
    \textbf{upon} receiving message \msg{transactions}{T, \sigma} \textbf{do} \`// $T$ is a set of ordered transactions; $\sigma$ is a signature \label{}\\
    \> \textbf{if} $\op{verify}(T, \sigma) = \true$ \textbf{then} \label{}\\
    \>\> \textbf{for} $\var{tx} \in T$ \textbf{do}  \label{}\\
    \>\>\> \op{execute}(\var{tx})  \label{}
  \end{numbertabbing}
  }
  \caption{Pseudocode of a smart contract.}
  \label{alg:smartcontract}
\end{algo*}

\begin{figure}[!ht]
  \centering
  \includegraphics[width=1.0\textwidth]{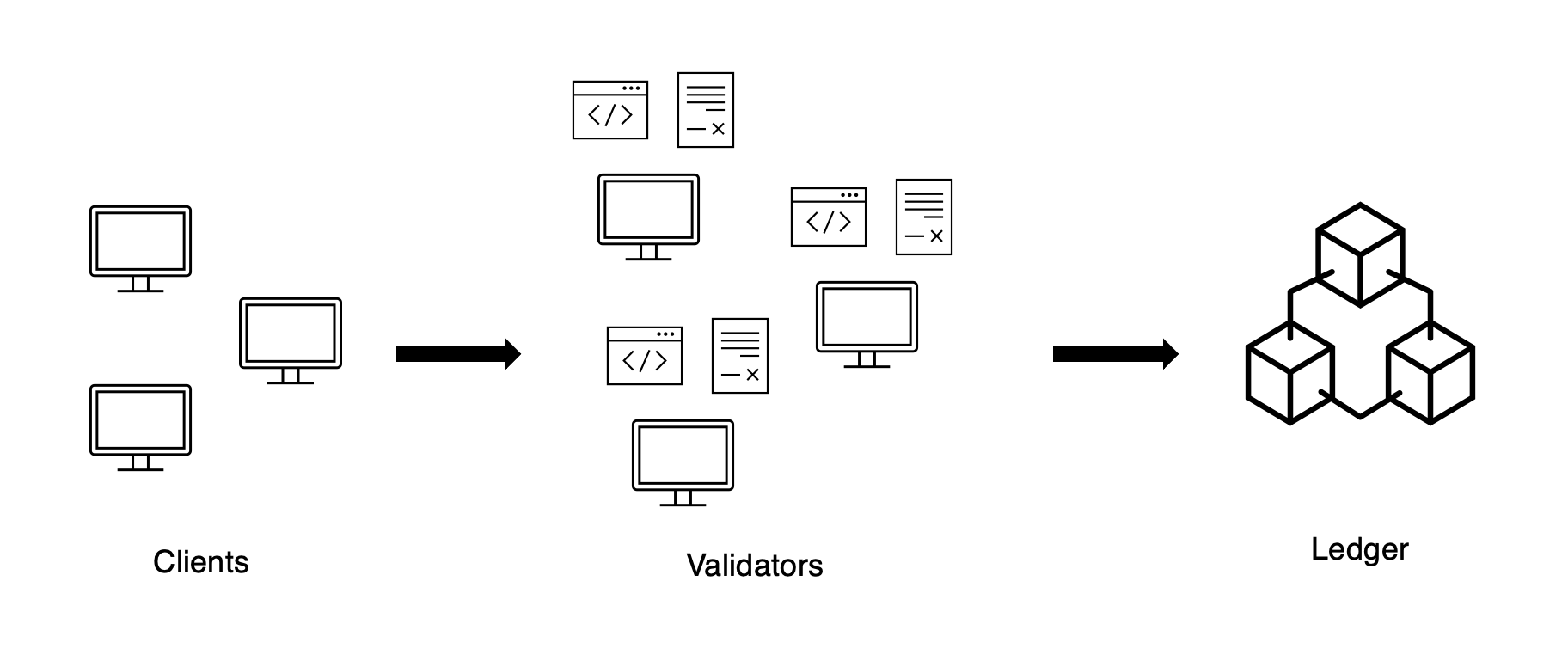}
  \caption{Clients submit transactions to the validators running the QOF algorithm. Validators agree on a fair order of transactions and run a smart contract to execute transactions on the ledger.}
  \label{fig:int2}
\end{figure}

\paragraph{Integration with consensus.}
The second approach is to directly build the QOF protocol into the consensus protocol run by the validators, as shown in Figure~\ref{fig:int2}. In this case, clients submit transactions to the validators that extend their consensus protocol by the QOF algorithm and output transactions in fair order. Then, validators run a smart contract to execute transactions on the ledger. This approach implements the fairness notion natively and for all smart contracts and realizes the original vision of protecting the system against front-running. A notable disadvantage is the requirement to change the original protocol since the QOF algorithm needs to be integrated. Furthermore, the very notion of differential order-fairness relies on a known set of validators. Hence, such an integration is not feasible for truly permissionless consensus protocols.

In particular, layer-one protocols like \emph{Tendermint Core}\footnote{\url{https://tendermint.com}} are suitable for integrating QOF protocol based on their trust model. Tendermint is a Byzantine Fault Tolerant (BFT) consensus protocol that tolerates up to one-third of failures by stake. Tendermint forms the basis of all blockchain networks in the \emph{Cosmos ecosystem}, a collection of interoperable networks that comes with tools for efficient and secure communication and coordination between the individual blockchains. There exist many other blockchain networks with stake-based consensus, into which the QOF protocol may be integrated (Algorand\footnote{\url{https://algorandtechnologies.com}}, Cardano\footnote{\url{https://cardano.org}}, Internet Computer/DFINITY\footnote{\url{https://internetcomputer.org}}, Avalanche\footnote{\url{https://www.avax.network}} and more).

\emph{Aptos}\footnote{\url{https://aptos.dev}} and
\emph{Sui}\footnote{\url{https://sui.io}}~\cite{suidocs} use related
consensus models, but differ in a crucial aspect.  Aptos runs
\emph{Block-STM}~\cite{DBLP:conf/ppopp/GelashviliSXDLM23}, an engine for
parallel execution for smart contracts, and Sui works in a permissionless
setting, where transactions are sent through a form of consistent broadcast
that does not establish consensus. However, both also use quorums at their core,
like the other protocols mentioned earlier. Since neither Aptos nor Sui
imposes a total order on all transactions, they permit some amount
of concurrent execution, which greatly improves scalability compared to
traditional Byzantine agreement methods. This poses a challenge for
integrating differential order fairness with their execution model, which
remains open at this time.

\section{Evaluation}
\label{sec:evaluation}
We evaluate the performance of the QOF protocol for estimating the cost of adding order fairness. Consensus is implemented by the HotStuff protocol in both cases. We first describe the experimental setup and then present the results.

\subsection{Experimental setup}
The starting point of our implementation of quick order fairness was the HotStuff implementation in the \emph{bamboo} project~\cite{DBLP:conf/icdcs/GaiFNFBD21}. The library\footnote{\url{https://github.com/gitferry/bamboo}} is implemented in the Go language and has several HotStuff flavors. Therefore, we modified the original code slightly to integrate it with the QOF protocol. Concretely, we modified the basic HotStuff protocol. The exact modifications are shown in the Section~\ref{sec:implementation}. 

As the baseline for the evaluation, we use \emph{bamboo}'s HotStuff implementation in the same benchmark setup as the QOF protocol. Therefore, we can compare the performance of both protocols in the same environment. We also compare and discuss the performance of QOF with other protocols such as Themis~\cite{DBLP:conf/ccs/KelkarDLJK23}, Pompē~\cite{DBLP:conf/osdi/ZhangSCZA20}, and unmodified bamboo HotStuff~\cite{DBLP:conf/icdcs/GaiFNFBD21}. Although these protocols are tested in different benchmark setups, we can still get a rough idea of how QOF performs compared to other protocols.

Our benchmarks measure and analyze the protocol's latency (in milliseconds) and throughput (transactions per second). An important question that arises is how to measure these metrics. Should we measure time from when the client submits a transaction (client latency) or when a party receives it (server latency)? We chose server latency because it is more relevant to the protocol's performance. The client would additionally depend on the network delay between the client and the server.

All benchmarks are made on one Linux virtual machine running Ubuntu 22.04 within an OpenStack hypervisor, with 32 GB memory and 16 vCPUs of an AMD EPYC-Rome Processor at 2.3GHz and 4500 bogomips. In our benchmark, we vary the number of servers from 4 to 64 to see how the protocol scales. We generate transactions from a separate client party and send them to all servers. 

\subsection{Results}
In this benchmark, we report the latency and throughput of the quick order fairness protocol and compare it to the baseline HotStuff protocol. We focus first on scalability and then evaluate how the payload size of a transaction and network delay affect the protocol's performance. Finally, we compare the performance of the QOF protocol to other protocols.

\begin{figure}[!ht]
  \centering
  \begin{subfigure}{.5\textwidth}
    \centering
    \includegraphics[scale=0.5]{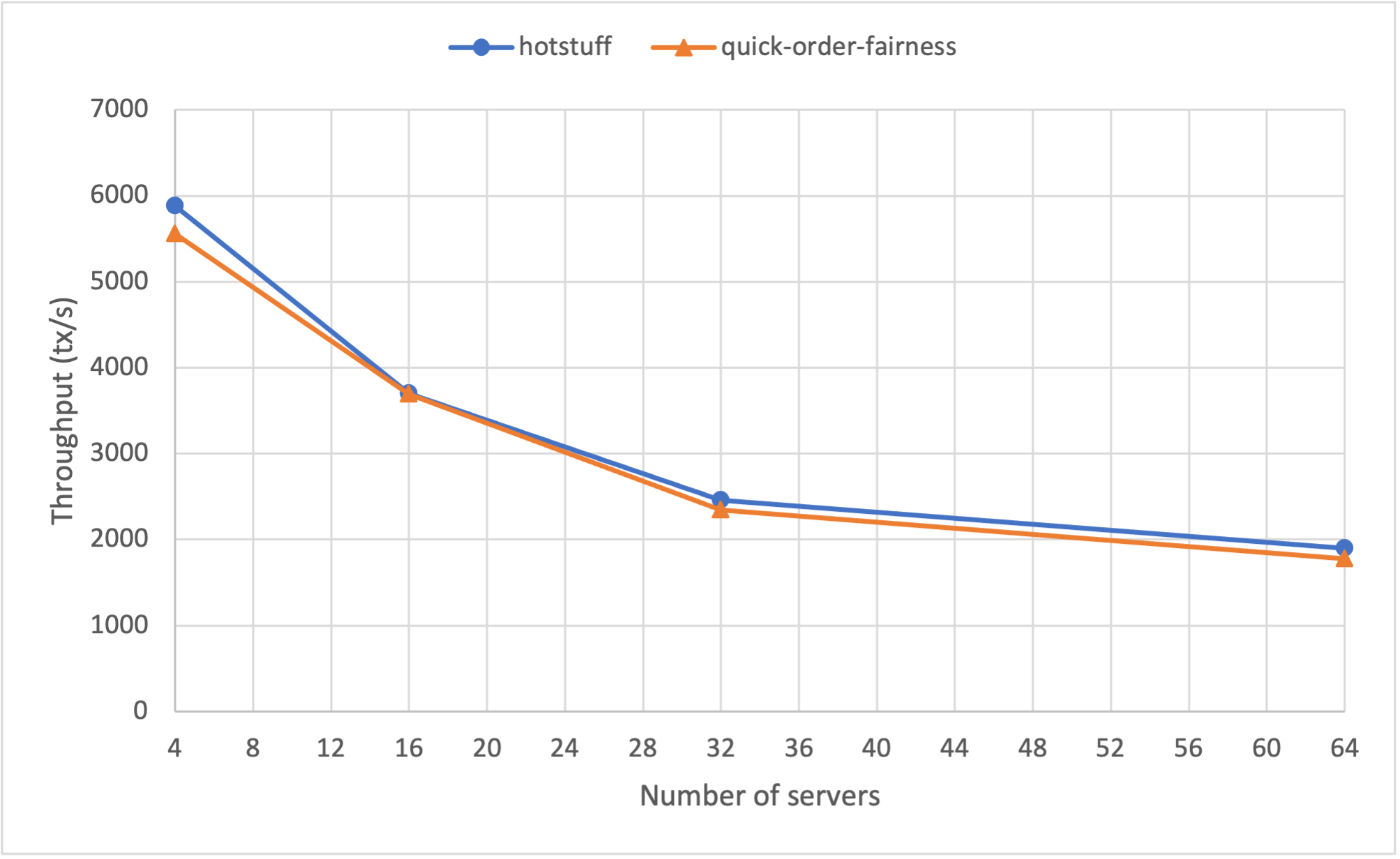}
    \caption{Throughput (tx/s)}
    \label{fig:throuhput}
  \end{subfigure}%
  \begin{subfigure}{.5\textwidth}
    \centering
    \includegraphics[scale=0.5]{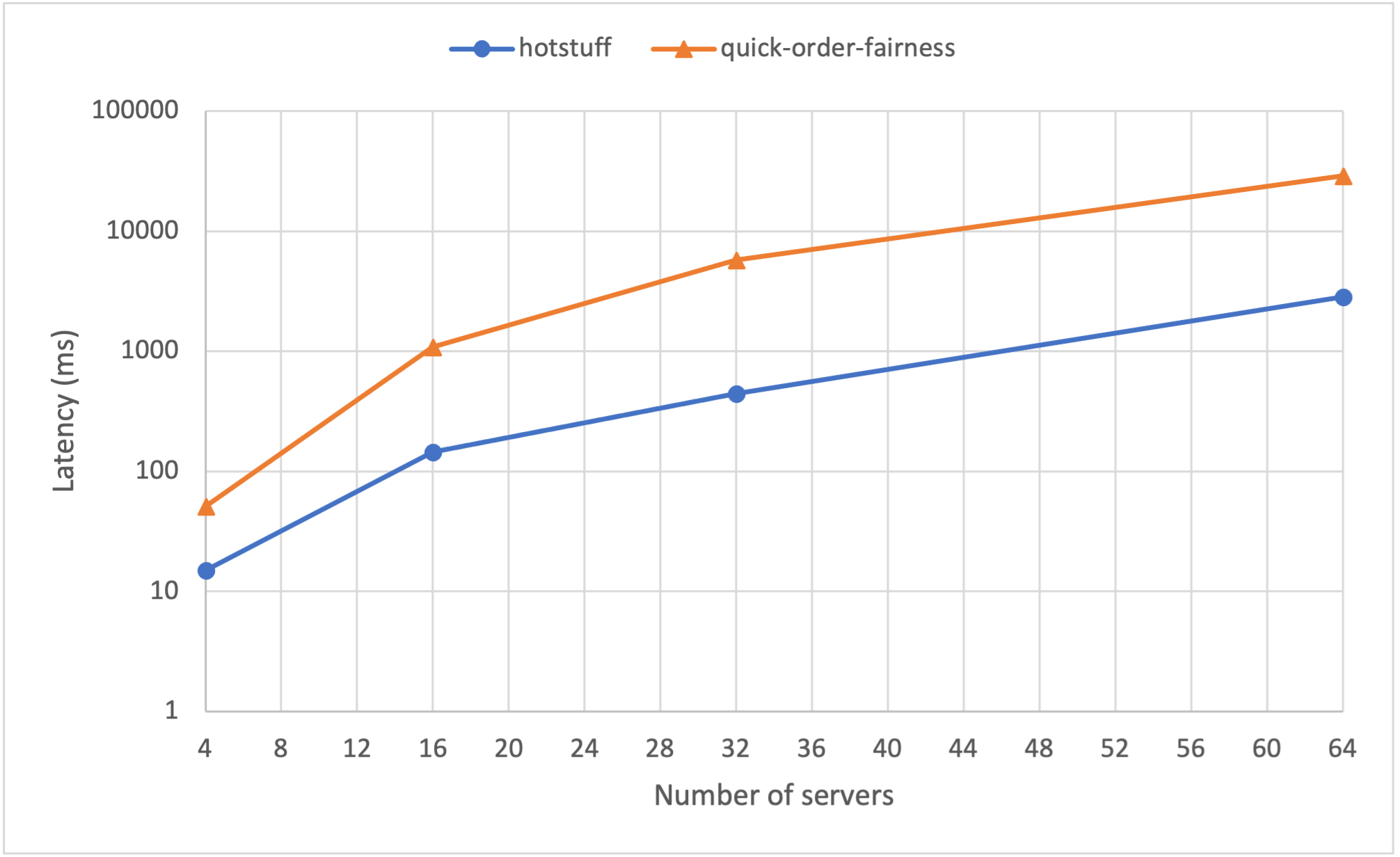}
    \caption{Latency (ms)}
    \label{fig:latency}
  \end{subfigure}
  \caption{Performance scalability of HotStuff and Quick Order Fairness protocols, showing how throughput degrades and latency increases with changing number of servers.}
  \label{fig:scalability}
  \end{figure}

\paragraph{Scalability.} A scalability benchmark is crucial for assessing a system's ability to handle increased workloads effectively. Figure~\ref{fig:throuhput} shows how the throughput in both protocols changes with the increase in the number of servers. With four servers, QOF reduces the throughput compared to HotStuff by about 300 transactions per second or approximately 5\%. Throughput reduction shrinks as the number of servers increases; with 64 servers, the reduction is about 6\%.

Figure~\ref{fig:latency} depicts the increase of latency with the increase of server numbers. Here, we observe that QOF increases latency by about 36 ms for four servers, which is around three times higher than HotStuff, and this difference continues to grow as the number of servers increases. The difference is primarily due to the increased computational load for processing the graph, which becomes more complex with more vertices and servers. Moreover, quick order fairness employs a Byzantine consistent broadcast channel that adds latency and computational cost compared to the HotStuff implementation.

\begin{figure}[!ht]
  \centering
  \begin{subfigure}{.5\textwidth}
    \centering
    \includegraphics[scale=0.5]{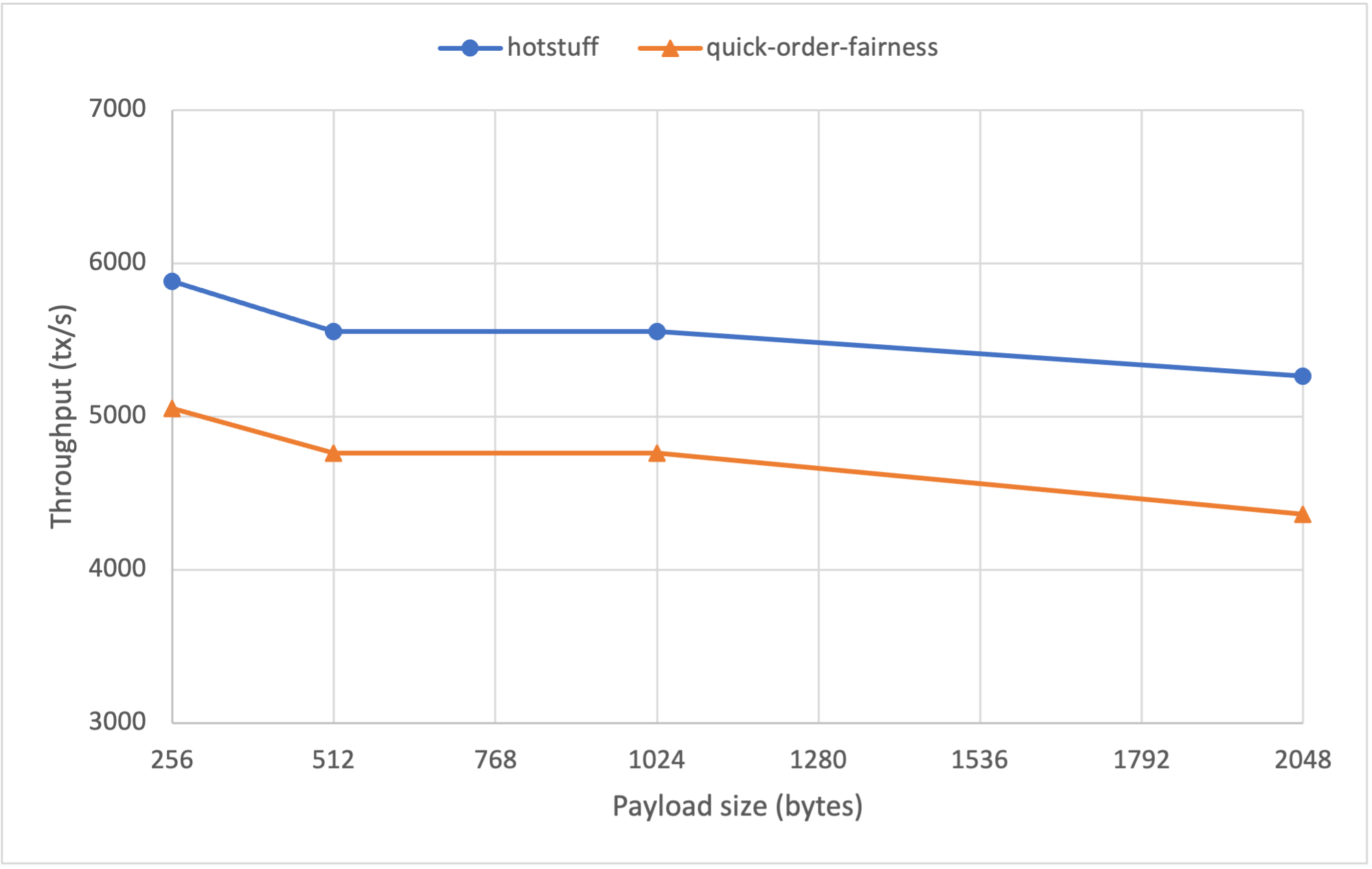}
    \caption{Throughput (tx/s)}
    \label{fig:payload-th}
  \end{subfigure}%
  \begin{subfigure}{.5\textwidth}
    \centering
    \includegraphics[scale=0.5]{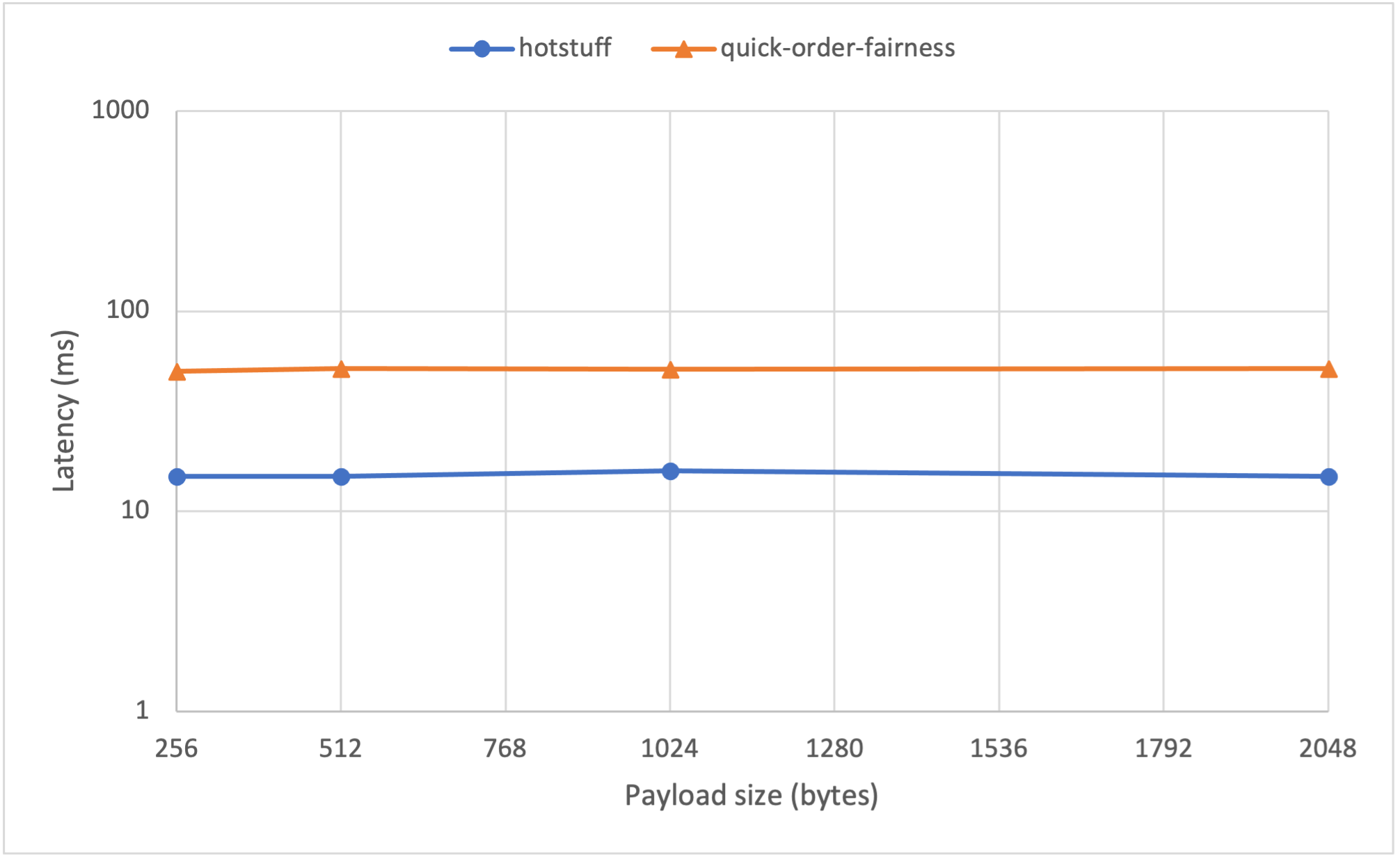}
    \caption{Latency (ms)}
    \label{fig:payload-lat}
  \end{subfigure}
  \caption{The figure shows how the throughput and latency of the quick order fairness and HotStuff protocol changes for different payload sizes. We vary the payload size from 246 to 2048 bytes.}
  \label{fig:payload}
  \end{figure}

\paragraph{Transaction payload size.}
In this benchmark, we analyze how the payload size of a transaction affects the performance of the quick order fairness protocol and HotStuff. We vary the payload size of a transaction from 256 bytes to 2048 bytes and show the results for the setup with four servers. Figure~\ref{fig:payload-th} shows the throughput of the protocol. The throughput data shows a gradual decrease as the payload size increases, with the throughput experiencing a reduction of approximately 14\%. Figure~\ref{fig:payload-lat} shows the protocol latency, and we see that the latency is constant for all payload sizes. We conclude that the payload size does not affect the performance of the quick order fairness and HotStuff protocol.

\begin{figure}[!ht]
  \centering
  \begin{subfigure}{.5\textwidth}
    \centering
    \includegraphics[scale=0.5]{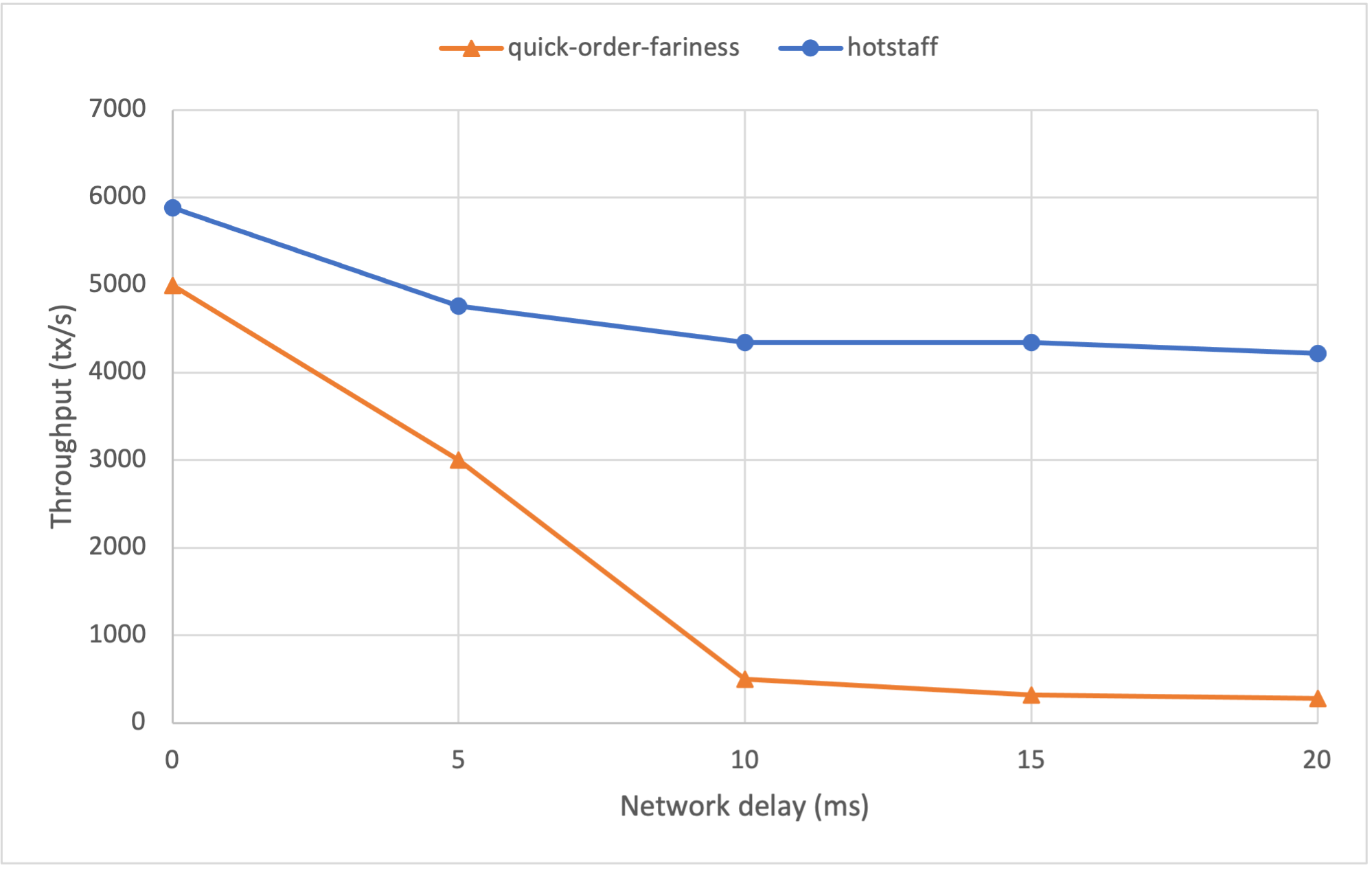}
    \caption{Throughput (tx/s)}
    \label{fig:net-th}
  \end{subfigure}%
  \begin{subfigure}{.5\textwidth}
    \centering
    \includegraphics[scale=0.5]{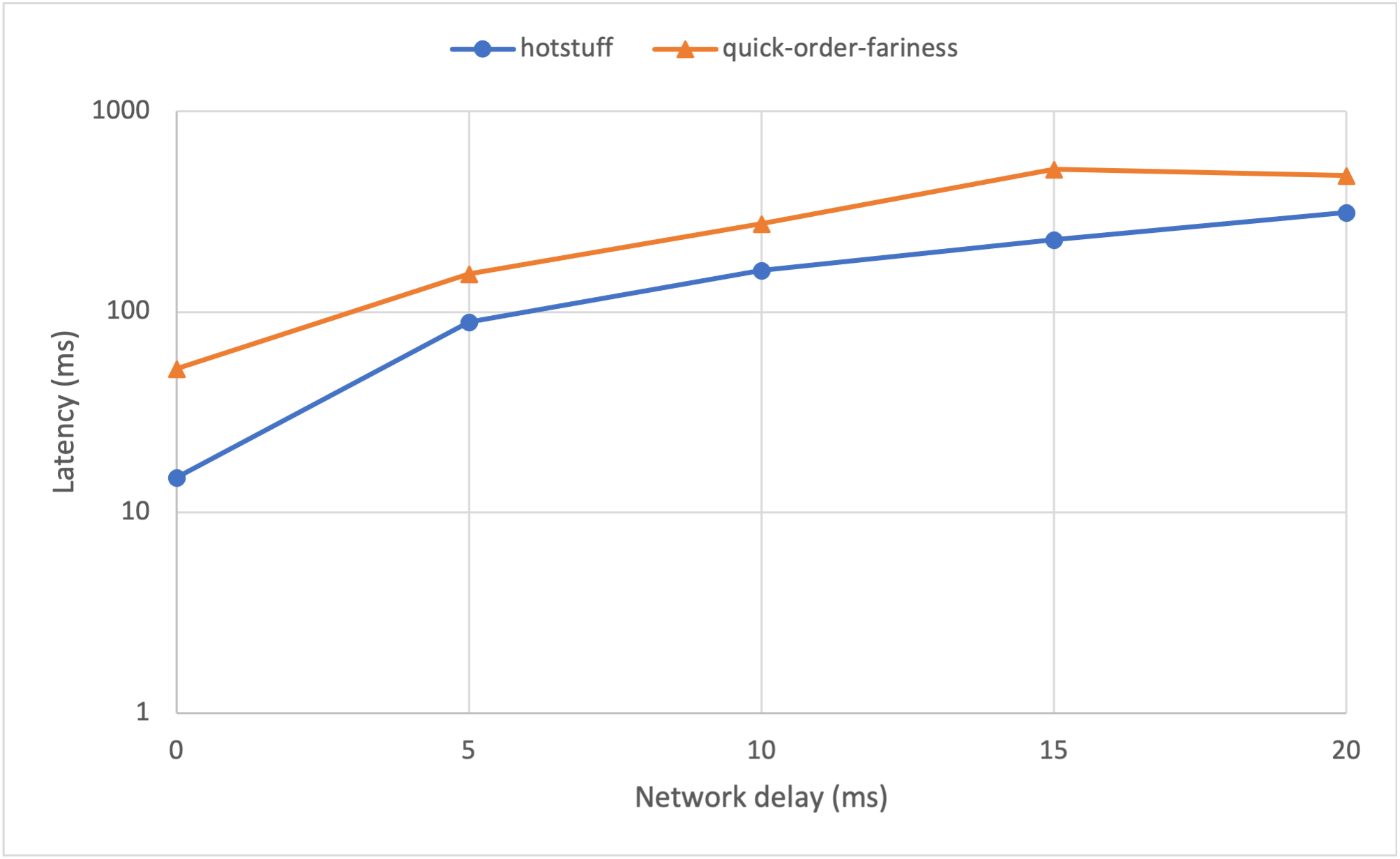}
    \caption{Latency (ms)}
    \label{fig:net-lat}
  \end{subfigure}
  \caption{The figure shows how the throughput and latency of the quick order fairness and HotStuff protocol changes for different network delays. The network delay changes from 0 to 20 milliseconds.}
  \label{fig:delay}
  \end{figure}

\paragraph{Network delay.} In this test, we introduce additional network delay as it might happen in the real-world environment. We vary the network delay from 0 to 20 milliseconds. Specifically, the delay is determined within a range defined by the configuration settings, introducing variability that mirrors the unpredictability of actual network latencies. This approach enhances the realism of the test environment, enabling a more comprehensive evaluation of the system's performance under diverse network delay scenarios. Measurements taken in a network with four servers are shown in Figures~\ref{fig:net-th} and \ref{fig:net-lat}, in terms of throughput and latency, respectively. The results indicate a substantial impact of network delay on system performance. As network delay increases from 0 to 20 ms, throughput decreases significantly while latency experiences a substantial increase. These findings underscore the sensitivity of throughput and latency to network delays, emphasizing the importance of minimizing network latency for optimal system performance.

\paragraph{Comparison to other works.} Comparing our benchmarks to other evaluations from the literature poses a challenge since the benchmark setups differ. Therefore, we can only get a rough idea of how the quick order fairness protocol performs compared to other protocols. In this comparison, we look at the performance of the Themis~\cite{DBLP:conf/ccs/KelkarDLJK23}, Pompē~\cite{DBLP:conf/osdi/ZhangSCZA20} HotStuff variant, and unmodified HotStuff implemented in bamboo~\cite{DBLP:conf/icdcs/GaiFNFBD21}. 

All benchmarks were conducted on diverse virtual machine providers, each instance equipped with 16 vCPUs and 32 GB of memory (except for Pompē, which utilized a machine with 64GB memory). We restrict our comparison to a common data point in all benchmark scenarios involving 32 servers executing the respective protocols.

Themis demonstrates remarkable performance, achieving a throughput 21 times higher than QOF. However, it is crucial to note that this substantial difference is influenced by the dedicated virtual machine per server in Themis, allowing much more parallelization. In contrast, QOF employs a setup where all 32 servers run within one VM, sharing the same vCPU.

The second-best performance is observed in the HotStuff variant of Pompē, surpassing QOF by approximately four times in terms of throughput, with a reduction in latency by a factor of three. This performance difference is presumably also influenced by the allocated dedicated VMs per server and by variations in the HotStuff implementation. Pompē leverages the original HotStuff implementation~\cite{libhotstuff} written in~C++.

Finally, we executed the HotStuff variant of bamboo within the same virtual machine as QOF. Specifically, in the scenario involving 32 servers, bamboo HotStuff achieves a throughput two times higher than QOF. These performance results for bamboo HotStuff can be attributed to the complex graphs constructed by QOF. Furthermore, introducing a Byzantine-consistent broadcast channel incurred additional latency and computational costs compared to the HotStuff implementation.

\section{Conclusion}
\label{sec:conclusion}
Through the practical implementation of the QOF protocol, the paper offers an executable representation, enhancing the protocol's accessibility and applicability. The systematic exploration of the protocol's integration into real-world systems, complete with smart contract pseudocode, lays a foundation to integrate the QOF protocol. The empirical evaluation, containing critical dimensions like scalability, throughput, and latency, not only confirms the protocol's efficacy but also provides invaluable insights for its practical deployment. Future work should optimize the codebase to enhance throughput and latency, ensuring the QOF protocol is ready for real-world deployment. Despite a slight reduction in throughput compared to the HotStuff protocol, the QOF protocol's complexity is justified by its resilience against front-running attacks. This work contributes practically and theoretically, extending the understanding and applying the quick order fairness protocol in decentralized systems.

\section*{Acknowledgments}
We thank Luca Zanolini for helping us implement the validated Byzantine consensus module and Ignacio Amores Sesar for participating in discussions. This work has been funded by the Swiss National Science Foundation (SNSF)
under grant agreement Nr\@.~200021\_188443 (Advanced Consensus Protocols) and from the Ripple University Blockchain Research Initiative.

\bibliography{references, dblpbibtex}
\bibliographystyle{ieeesort}

\end{document}